\documentclass[aps,prb]{revtex4}
\usepackage{booktabs}
\usepackage{textgreek}
\usepackage{threeparttable}
\usepackage{bm}
\usepackage{amsmath}
\usepackage{color}
\usepackage{graphicx}
\usepackage{amssymb}
\usepackage{soul}
\usepackage{tablefootnote}
\DeclareMathAlphabet{\mathbbmsl}{U}{bbm}{m}{sl}
\makeatletter
\newsavebox{\@brx}
\newcommand{\llangle}[1][]{\savebox{\@brx}{\(\m@th{#1\langle}\)}%
\mathopen{\copy\@brx\kern-0.5\wd\@brx\usebox{\@brx}}}
\newcommand{\rrangle}[1][]{\savebox{\@brx}{\(\m@th{#1\rangle}\)}%
\mathclose{\copy\@brx\kern-0.5\wd\@brx\usebox{\@brx}}}
\makeatother

\begin{document}
\draft

\title{Rashba spin-orbit coupling and quantum-interference effect for a pair of spin-correlated electrons in their tunneling and reflection under a step potential}

\author{Po-Hsin Shih$^{1}$, Godfrey Gumbs$^{1}$, Danhong Huang$^{2}$\footnote{E-mail contact: danhong.huang@spaceforce.mil}, Andrii Iurov$^{3}$ and Yonatan Abranyos$^{1}$}
\address{Department of Physics, Hunter College, City University of New York, 695 Park Avenue, New York, NY 10065 USA\\
$^{2}$US Air Force Research Laboratory, Space Vehicles Directorate (AFRL$/$RVSU),\\
Kirtland Air Force Base, New Mexico 87117, USA\\
$^{3}$Department of Physics and Computer Science, Medgar Evers College of City University of New York, Brooklyn, NY 11225, USA
}

\date{\today}

\begin{abstract}
We present both theory and numerical-computation results for the transmission and reflection probability currents of a charged particle across a potential step in the presence of 
a Rashba spin-orbit interaction. By varying kinetic energy and angle of incident electrons or barrier height, different features associated with tunneling and reflection of electrons 
are revealed by inter-spin-channel electron tunnelings and reflections. These unique properties are further accompanied by spin-state quantum interference of either a reflected or 
transmitted pairs of spin-correlated electrons with the same kinetic energy but in different spin-orbital states. Such distinctive features are expected to give rise to a lot of applications 
in both spintronics and quantum-computation devices.
\end{abstract}

\maketitle

\section{Introduction}
\label{sec1}

It is well known that the spin-orbit interaction (SOI) is essentially a relativistic effect, i.e., a moving electron under an external electric field experiences a magnetic field in its rest frame.\,\cite{book1} In a semiconductor, the SOI enables an electron undergoing a spin procession while it passes through the material, which constitutes a basis of various proposed semiconductor-based spintronic devices.\,\cite{add1,add2,add3} 
In nanostructures, quantum confinement can further modify the symmetry of SOI.\,\cite{add4} From physics perspective, the relativistic motion of an electron can be described by a Dirac equation. By combining these effects, one expects both an electric dipole moment as well as Thomas procession\,\cite{thomas} due to the
rotational kinetic energy under an electric field.\,\cite{new6,new7} Mathematically, however, the SOI Hamiltonian can be derived from the Dirac equation after taking the non-relativistic limit up to terms of $\sim(v/c)^2$, where $v$ is the velocity of an electron. Formally, this limit can be reached by either expanding the Dirac equation in powers of $\sim(v/c)$ or making use of the asymptotically exact Foldy-Wouthousen transformation.\,\cite{new8}
\medskip

In recent years, there have been a lot of interests in the research field dealing with the effect of SOI or the Rashba effect, on the electron transport and optical properties of low-dimensional semiconductor electronic systems.\,\cite{new1,new2,new3,new4,new19,new20,new21,czhang} Such an SOI results from the asymmetry in a confining potential for an electron or a hole gas at the interface of a hetero-structure. Particularly, some of these studies aim at identifying potential spintronic-device applications, e.g., a spin-based transistor, in which the spin-current can be manipulated purely by electrical means (i.e., spin-Hall effect)\,\cite{new5,add5,add6}. Meanwhile, special focus has also been put on its effect on plasmon excitations.\,\cite{new9,new10,new11,new12,new13,new14,new15,new16,new17,new18}
\medskip

Very recently, by utilizing controlled polarization of incident light and soft-X-ray angle-resolved photoemission spectroscopy, the giant Rashba effect and spin polarization in GeTe (a component of chalcogenide glasses) thin films as functions of light polarization and angle of emission were revealed and analyzed systematically.\,\cite{add6} For non-ferromagnetic GeTe with a giant Rashba parameter, one expects that a Zeeman gap will open at the Dirac point under a magnetic field.\,\cite{vijay,add7} In general, the Fermi
surface of spin-orbit coupled materials consists of two concentric, spin non-degenerate hyper-surfaces separated in momentum space.\,\cite{new24} 
In the specific case of Rashba SOI materials, these two Rashba bands have opposite chiralities with the angle
between the spin and momentum vectors locked and reversed with respect to each other.
\medskip

In this paper, base on our calculated energy eigenstate for Rashba-Zeeman coupled two-dimensional conduction electrons in the presence of a spin-split gap, we obtain probability currents explicitly 
for both reflection and transmission of electrons under a step-like potential barrier. By matching boundary conditions at the interface of this step barrier, we acquire both spin-dependent reflection and transmission coefficients as functions of incident kinetic energy and angle of incidence, as well as of barrier height and steepness, for various spin directions, Zeeman gaps and Rashba parameter values. Our current study demonstrates the existence of paired Rashba-Zeeman-coupled inter-spin-channel tunneling of electrons with different diffraction angles for possible spin-state quantum interference in the system. Such a unique and attractive property can be utilized for designing non-magnetic spintronics and quantum-computation devices. 
\medskip

The rest of paper is organized as follows. In Sec.\,\ref{sec2}, we establish a theory for calculating exactly energy eigenstates of two-dimensional Rashba-Zeeman-coupled electrons with a spin-split gap. Moreover, the explicit expressions for probability currents, as well as for transmission and reflection coefficients, are derived in Sec.\,\ref{sec3}, corresponding to in-plane Rashba-Zeeman-coupled electron tunneling into a step-potential barrier. Numerical results, and their analysis and discussions are given in Sec.\,\ref{sec4} for tunneling and reflection coefficients as functions of incident kinetic energy, angle of incidence, barrier height, and barrier steepness 
at the interface. Finally, a brief summary is given in Sec.\,\ref{sec5}.

\section{Energy Eigen-States with a Gap}
\label{sec2}

Let us first consider a two-dimensional electron-gas system in $x$-$y$ plane with both Rashba spin-orbit and Zeeman couplings\,\cite{10} for orbital motions of electrons. For this model system, the Hamiltonian operator $\hat{{\cal H}}$ can be written as 

\begin{equation}
\hat{{\cal H}}=\left[\begin{array}{cc}
\alpha\,(\hat{p}_x^2+\hat{p}_y^2)  +V_B(x)-\Delta_Z\ \ & \alpha_R\left(\hat{p}_y- i\hat{p}_x\right)\\
\\
\alpha_R\left(\hat{p}_y+ i\hat{p}_x\right)\ \ & \alpha\,(\hat{p}_x^2+\hat{p}_y^2) +V_B(x)+\Delta_Z
\end{array}\right]\ ,
\label{e1}
\end{equation}
where $\hbar\alpha_R$ stands for the so-called Rashba parameter, $2\Delta_Z$ is the Zeeman gap produced by a localized in-plane magnetic field,
$\alpha=1/2m^*$ with an electron effective mass $m^*$ (positive for $n$-doping but negative for $p$-doping), 
$\hat{p}_x=-i\hbar\,\partial/\partial x$ and $\hat{p}_y=-i\hbar\,\partial/\partial y$ are two momentum operators in the $x$ and $y$ directions,  and $V_B(x)=V_0\,\Theta(x) $ represents a step potential with a barrier height $V_0>0$  along the $x$ direction. Consequently, the translation symmetry (or the electron transverse wave number $k_y$) along the $y$ direction retains for this system.
\medskip

We assume that the trial spinor-type wave function $\Psi(\mbox{\boldmath$r$})$ for our model system takes the form

\begin{equation}
\Psi(\mbox{\boldmath$r$})=\frac{\texttt{e}^{ik_yy}}{\sqrt{{\cal L}_y}}\,
\left[\begin{array}{c}
\psi_A(x)\\
\\
\psi_B(x)
\end{array}\right]\ ,
\label{e2}
\end{equation}
where ${\cal L}_y$ stands for the sample size in the $y$ direction and the subscripts $A,\,B$ label two components of a spinor-type wave function for the up- and down-pseudo-spin states of electrons.  From the static Schr\"odinger equation $\hat{{\cal H}}\,\Psi(\mbox{\boldmath$r$})=\overline{E}\,\Psi(\mbox{\boldmath$r$})$, we arrive at the following pair of homogeneous eigenvalue equations

\begin{equation}
\left[\begin{array}{cc}
-(\hbar^2\alpha)\,\partial^2/\partial x^2+V_B(x)+\hbar^2\alpha  k_y^2-\Delta_Z -\overline{E}\ \ \ & \alpha_R\hbar\left(k_y -   \partial/\partial x \right)\\
\\
\alpha_R\hbar\left(k_y +  \partial/\partial x \right)\ \ \ & -(\hbar^2 \alpha)\,\partial^2/\partial x^2+V_B(x)+\hbar^2\alpha  k_y^2 +\Delta_Z -\overline{E}
\end{array}\right]
\left[\begin{array}{c}
\psi_A(x)\\
\\
\psi_B(x)
\end{array}\right]=0\ ,
\label{e3}
\end{equation}
where $\overline{E}$ is the eigen-energy of electrons to be determined.  Mathematically, for such a second-order differential equation, we require two boundary conditions to select out a specific  solution for our considered system.
\medskip

In the special case of $V_B(x)=0$, we find $\hat{{\cal H}}\to\hat{{\cal H}}_0$, where $\hat{{\cal H}}_0$ can be simply obtained from Eq.\,\eqref{e1} by the substitution: $\partial/\partial x\to ik_x$. As a result, Eq.\,\eqref{e3} directly leads to a pair of spin-split bands, i.e.,

\begin{equation}
E_s(k)\equiv\overline{E}_s(k)-E_{\rm min}=\alpha \hbar^2k^2   +s\sqrt{\alpha^2_R\hbar^2k^2+\Delta_Z^2}-E_{\rm min}=E_R\bar{k}^2+s\sqrt{4E_R^2\bar{k}^2+\Delta_Z^2}-E_{\rm min}\ .
%=  \hbar^2\alpha\left(k \pm k_R\right)^2-E_R\ ,
\label{e4}
\end{equation}
In Eq.\,\eqref{e4}, $\overline{E}_s(k)$ reaches a minimum $E_{\rm min}=-E_R(1+\Delta_Z^2/4E^2_R)<0$ at $k=k_{\rm min}\equiv\pm k_R\sqrt{1-\Delta_Z^2/4E_R^2}$,  
where $s=\pm 1$ specifies the split upper $(+)$ or lower $(-)$ branch respectively, and the zero-energy point is selected at the minimum of $E_s(k)$ or the middle point of Zeeman gap of $\overline{E}_s(k)$,
$\bar{k}=k/k_R$, $E_R=\alpha_R^2/4\alpha$, and $k_R=\alpha_R/2\hbar\alpha$. The inclusion of $E_{\rm min}$ in Eq.\,\eqref{e4} ensures that $E_s(k)\geq 0$, and it 
formally resembles a split ``left-shifted'' ($s=+1$, spin-up) or ``right-shifted'' ($s=-1$, spin-down) energy parabola as $\Delta_Z=0$, where $k=\sqrt{k_x^2+k_y^2}$. This phenomenon leads to the 
so-called {\em spin-Hall effect}\,\cite{new5,add5,add6} in the absence of an external magnetic field. 
The positive (negative) sign of $\alpha$ value corresponds to an upward or electron-like (a downward or hole-like) parabola, resulting in opposite signs for group velocities.  
At $k=0$, however, these two spin-split bands are mixed with each other if $\Delta_Z\neq 0$, and therefore, the spin-conservation requirement can be relaxed in this case. Here, whenever $\Delta_Z\neq 0$, 
$s$ can still be regarded as a branch index 
with $s=+1$ for the split-upper branch while $s=-1$ for the split-lower branch. Furthermore, two orthonormal wave functions, associated with $s=\pm 1$ branches in Eq.\,\eqref{e4}, are found to be

\begin{equation}
\Psi_{s,{\bf k}}^{\rm (i)}(\mbox{\boldmath$r$})=\frac{\texttt{e}^{ik_xx+ik_yy}}{\sqrt{{\cal A}}}
\left[\begin{array}{c}
 -is\gamma_s(k)\,\texttt{e}^{i\theta_{\bf k}}\\
\\
1
\end{array}\right]\,\frac{1}{\sqrt{1+\gamma_s^2(k)}}\ ,
\label{e5}
\end{equation}
where $\gamma_s(k)=\alpha_R\hbar k/(\sqrt{\alpha_R^2\hbar^2k^2+\Delta_Z^2}+s\Delta_Z)$ and it reduces to unity for $\Delta_Z\to 0$, 
$\theta_{\bf k}=\tan^{-1}(k_y/k_x)$ represents a phase angle, and ${\cal A}$ stands for the surface area of the sample.
\medskip

If we only focus on the electron states close to $k=0$, by neglecting the higher-order terms proportional to $\sim k^2$ in Eq.\,\eqref{e4}, we are left with $E_{\pm}(k)\approx\pm\hbar\alpha_Rk$ for $\Delta_Z=0$, which  approximately resemble the upper ($+$) cone for all allowed electron states and the lower ($-$) cone for excluding disallowed electron states, including both pseudo-spin states of electrons.  However, in this paper, we would always use the full expressions in Eq.\,\eqref{e4} for split energy dispersions in our calculations.

\section{Probability current and transmission coefficient}
\label{sec3}

We now turn attention  to an important part of our theory, i.e., the calculation of probability current as a first step in determining the transmission or reflection coefficient. 
In general, the time-dependent Schr{\"o}dinger equation for the Hamiltoniam  $\hat{{\cal H}}$ in Eq.\,\eqref{e1}, as well as the spinor-type wave-function $|\mbox{\boldmath$\Psi$}>=[\psi_A\ \psi_B]^T$, can be written as $i\hbar\,\partial/\partial t\,|\mbox{\boldmath$\Psi$}>=\hat{{\cal H}}\,|\mbox{\boldmath$\Psi$}>$ from which we deduce that the probability density $\rho=<\mbox{\boldmath$\Psi$}|\mbox{\boldmath$\Psi$}>$ satisfies

\begin{equation}
i\hbar\frac{\partial\rho}{\partial t} =i\hbar\frac{\partial}{\partial t} <\mbox{\boldmath$\Psi$}|\mbox{\boldmath$\Psi$}>
= <\mbox{\boldmath$\Psi$}|\hat{{\cal H}}|\mbox{\boldmath$\Psi$}>
-\left(   
 \hat{{\cal H}}|\mbox{\boldmath$\Psi$}>\right)^\dag |\mbox{\boldmath$\Psi$}>  \   .
\label{GG1}
\end{equation}
Inserting the Hamiltonian in Eq.\,\eqref{e1} into Eq.\,\eqref{GG1}, we find, after a lengthy calculation, that the probability-current density $\mbox{\boldmath$j$}=\{j_x,\,j_y\}$ satisfies the continuity equation  $\nabla\cdot\mbox{\boldmath$j$}+\partial\rho/\partial t =0$, and its two components, $j_x$ and $j_y$, are found to be

\begin{subequations}
\begin{align}
\label{GG2x}
j_x&= 2\alpha\hbar\sum_{\nu=A,B} {\rm Im}\left( \psi_\nu^\ast  \frac{\partial\psi_\nu}{\partial x} \right)
+2\alpha_R\,{\rm Im}\left( \psi_A^\ast\psi_B \right)\ ,\\
\label{GG2y}
j_y&= 2\alpha\hbar\sum_{\nu=A,B} {\rm Im}\left( \psi_\nu^\ast  \frac{\partial\psi_\nu}{\partial y} \right)
+2\alpha_R\,{\rm Re}\left( \psi_A^\ast\psi_B \right) \  ,
\end{align}
\end{subequations}
where the first term on the right-hand-sides relates to the pure orbital current while the second term results from the spin-orbital-coupling current. Substituting the wave function in Eq.\,(\ref{e5}) into 
Eqs.\,(\ref{GG2x}) and \eqref{GG2y}, we obtain the incident probability current $\mbox{\boldmath$J$}^{(\rm i)}(s,\mbox{\boldmath$k$})=\{J_x^{\rm (i)}(s,\mbox{\boldmath$k$}),\,J_y^{\rm (i)}(s,\mbox{\boldmath$k$})\}$ 
for $s=\pm 1$

\begin{subequations}
\begin{align}
\label{GG3x}
J_x^{\rm (i)}(s,\mbox{\boldmath$k$}) &=  j_x^{\rm (i)}(s,\mbox{\boldmath$k$}){\cal A}=2\left[s\Gamma_s(k)\alpha_R\cos\theta_{\bf k} +\alpha\hbar k_x\right]
\equiv I_s(k)\cos\theta_{\bf k}\  ,\\
\label{GG3y}
J_y^{\rm (i)}(s,\mbox{\boldmath$k$}) &=  j_y^{\rm (i)}(s,\mbox{\boldmath$k$}){\cal A}=2\left[s\Gamma_s(k)\alpha_R\sin\theta_{\bf k} +\alpha\hbar k_y\right]
\equiv I_s(k)\sin\theta_{\bf k}\  ,
\end{align}
\end{subequations}
where $\Gamma_s(k)=\gamma_s(k)/[1+\gamma_s^2(k)]\to 1/2$ for $\Delta_Z=0$, 
$I_s(k)=2[s\Gamma_s(k)\alpha_R +\alpha\hbar k]$ and $\{k_x,k_y\}=\{k\cos\theta_{\bf k},\,k\sin\theta_{\bf k}\}$.
In particular, for an incident particle, we should designate both its branch index $s$ and kinetic energy $E_s(k)$ in Eq.\,\eqref{e4} simultaneously.
Similarly, the probability current $\mbox{\boldmath$J$}^{(\rm t)}(s'',\mbox{\boldmath$q$};\,|t|^2)=\{J_x^{\rm (t)}(s'',\mbox{\boldmath$q$};\,|t|^2),\,J_y^{\rm (t)}(s'',\mbox{\boldmath$q$};\,|t|^2)\}$ for the transmitted wave with branch index $s^{\prime\prime}=\pm 1$ and the wave function given by 

\begin{equation}
\Psi_{s^{\prime\prime},{\bf q}}^{\rm (t)}(\mbox{\boldmath$r$})=t\  \frac{\texttt{e}^{iq_xx+iq_yy}}{\sqrt{{\cal A}}}
\left[\begin{array}{c}
 -is^{\prime\prime}\gamma_{s^{\prime\prime}}(q)\,\texttt{e}^{i\theta^{\prime\prime}_{\bf q}}\\
\\
1
\end{array}\right]\,\frac{1}{\sqrt{1+\gamma_{s^{\prime\prime}}^2(q)}}\ ,
\label{e5t}
\end{equation}
is calculated as

\begin{subequations}
\begin{align}
\label{GG3tx}
J_x^{\rm (t)}(s^{\prime\prime},\mbox{\boldmath$q$};\,|t|^2) &=j_x^{\rm (t)}(s^{\prime\prime},\mbox{\boldmath$q$};\,|t|^2){\cal A}= 2|t|^2\left[s^{\prime\prime}\Gamma_{s^{\prime\prime}}(q)\alpha_R\cos\theta^{\prime\prime}_{\bf q}+\alpha^{\prime\prime}\hbar q_x\right]   =|t|^2I_{s^{\prime\prime}}(q)\cos\theta^{\prime\prime}_{\bf q}\  ,\\
\label{GG3ty}
J_y^{\rm (t)}(s^{\prime\prime},\mbox{\boldmath$q$};\,|t|^2) &=j_y^{\rm (t)}(s^{\prime\prime},\mbox{\boldmath$q$};\,|t|^2){\cal A}= 2|t|^2\left[s^{\prime\prime}\Gamma_{s^{\prime\prime}}(q)\alpha_R\sin\theta^{\prime\prime}_{\bf q}+\alpha^{\prime\prime}\hbar q_y\right]=|t|^2I_{s^{\prime\prime}}(q)\sin\theta^{\prime\prime}_{\bf q}\  ,
\end{align}
\end{subequations}
where $\theta^{\prime\prime}_{\bf q}=\tan^{-1}(q_y/q_x)=\tan^{-1}(k_y/q_x)$, $I_{s^{\prime\prime}}(q)=2[s^{\prime\prime}\Gamma_{s^{\prime\prime}}(q)\alpha_R+\alpha^{\prime\prime}\hbar q]$.  
In this notation, $\alpha^{\prime\prime}=\pm\alpha$, $q_x=q\cos\theta^{\prime\prime}_{\bf q}$ and $q_y=q\sin\theta^{\prime\prime}_{\bf q}=k_y=k\sin\theta_{\bf k}$, where the spin ($\alpha_R$) and the orbital ($\alpha$) currents can have opposite signs with respect to each other. Especially, for transmission with $\alpha_R=0$, 
we find $I_{s^{\prime\prime}}(q)\sin\theta^{\prime\prime}_{\bf q}=(\alpha^{\prime\prime}/\alpha)\,I_s(k)\sin\theta_{\bf k}$ for the conservation of probability current along the $y$ direction under $|t|^2=1$, 
which is formally similar to the well-known Snell's law after taking $\alpha^{\prime\prime}=\alpha$. However, the spin-orbit coupling still enables breaking down this conservation of probability current along the $y$ direction even for $s=s^{\prime\prime}$ since $q\neq k$ as $V_0\neq 0$. 
Here, the branch index $s^{\prime\prime}$ should be assigned in advance for either an inter-spin-channel ($s\neq s^{\prime\prime}$) or an intra-spin-channel ($s=s^{\prime\prime}$) tunneling of electrons. 
For given branch indexes $s,\,s^{\prime\prime}$, incident energy $E_s(k)=E_K$, incident angle $\theta_{\bf k}$, barrier height $V_0$, and the bandgap $2\Delta_Z$, the longitudinal wave number $q_x$ on the barrier side  
can be found from Eq.\eqref{kq2} below. Meanwhile, for the reflected wave with its wave function written as

\begin{equation}
\Psi_{s^\prime,{\bf k}^\prime}^{\rm (r)}(\mbox{\boldmath$r$})=r  \  \frac{\texttt{e}^{ik_yy -ik^{\prime}_xx}}{\sqrt{{\cal A}}}
\left[\begin{array}{c}
 -is^{\prime}\gamma_{s^{\prime}}(k^\prime)\,\texttt{e}^{i(\pi-\theta_{{\bf k}^\prime})}\\
\\
1
\end{array}\right]\,\frac{1}{\sqrt{1+\gamma_{s^{\prime}}^2(k^{\prime})}}\ ,
\label{e5r}
\end{equation}
the probability current $\mbox{\boldmath$J$}^{(\rm r)}(s',\mbox{\boldmath$k$}';\,|r|^2)=\{J_x^{\rm (r)}(s',\mbox{\boldmath$k$}';\,|r|^2),\,J_y^{\rm (r)}(s',\mbox{\boldmath$k$}';\,|r|^2)\}$ 
for the branch index $s^\prime$ is found to be 

\begin{subequations}
\begin{align}
\label{GG3rx}
J_x^{\rm (r)}(s^\prime,\mbox{\boldmath$k$}^\prime;\,|r|^2) &=j_x^{\rm (r)}(s^\prime,\mbox{\boldmath$k$}^\prime;\,|r|^2){\cal A}= -2|r|^2\left[s^\prime\Gamma_{s^\prime}(k^\prime)\alpha_R\cos\theta_{{\bf k}^\prime}+\alpha\hbar k^\prime_x\right]=  -   |r|^2I_{s'}(k^\prime)\cos\theta_{{\bf k}^\prime}\  ,\\
\label{GG3ry}
J_y^{\rm (r)}(s^\prime,\mbox{\boldmath$k$}^\prime;\,|r|^2) &=j_y^{\rm (r)}(s^\prime,\mbox{\boldmath$k$}^\prime;\,|r|^2){\cal A}= +2|r|^2\left[s^\prime\Gamma_{s^\prime}(k^\prime)\alpha_R\sin\theta_{{\bf k}^\prime}+\alpha\hbar k_y\right]=+|r|^2I_{s^\prime}(k^\prime)\sin\theta_{{\bf k}^\prime}\ ,
\end{align}
\end{subequations}
where $I_{s^{\prime}}(k^\prime)=2[s^{\prime}\Gamma_{s^{\prime}}(k^\prime)\alpha_R+\alpha\hbar k^\prime]$, $k^\prime_x=k^\prime\cos\theta_{{\bf k}^\prime}$, 
$k_y=k^\prime\sin\theta_{{\bf k}^\prime}=k\sin\theta_{\bf k}$, and the probability current along the $y$ direction is not always conserved, if $s^\prime\neq s$ or $k^\prime\neq k$, even for reflection under $|r|^2=1$. 
\medskip

Physically, we can further introduce a set of boundary conditions for our model system. 
Since the connection between two components $\psi_A(x)$ and $\psi_B(x)$ of a spinor-type wave function in Eq.\,\eqref{e2} have already been built into eigenvalue equation presented in Eq.\,\eqref{e3}, we need only consider one component, e.g., $\psi_A(x)$ 
(in the sense that the obtained $|t|^2$ and $|r|^2$ remain the same). 
\medskip

First, for given branch indexes $s,\,s^{\prime\prime}$, by using continuities for both wave functions and their derivatives on both sides of the step boundary at $x=0$, 
we acquire the following relations for unknown $r$ and $t$, yielding

\begin{subequations}
\begin{align}
\label{boundary2}
-is^{\prime\prime}\texttt{e}^{i\theta^{\prime\prime}_{\bf q}}\Gamma_{s^{\prime\prime}}(q)\,t&=-is\Gamma_s(k)\,\texttt{e}^{i\theta_{\bf k}}+is^\prime\Gamma_{s^\prime}(k^\prime)\,\texttt{e}^{-i\theta_{{\bf k}^\prime}}\,r\ ,\\
\label{boundary2p}
q_xs^{\prime\prime}\texttt{e}^{i\theta^{\prime\prime}_{\bf q}}\Gamma_{s^{\prime\prime}}(q)\,t&=k_xs\Gamma_s(k)\,\texttt{e}^{i\theta_{\bf k}}+k^\prime_xs^\prime\Gamma_{s^\prime}(k^\prime)\,\texttt{e}^{-i\theta_{{\bf k}^\prime}}\,r\ ,
\end{align}
\end{subequations}
which leads to the conclusion

\begin{subequations}
\begin{align}
\label{boundary3}
t(k_x,k^\prime_x,q_x\,\vert s,s^\prime,s^{\prime\prime})&=\left[\frac{s\Gamma_s(k)}{s^{\prime\prime}\Gamma_{s^{\prime\prime}}(q)}\right]\,\texttt{e}^{i(\theta_{\bf k}-\theta^{\prime\prime}_{\bf q})}\left[\frac{k^\prime\cos\theta_{{\bf k}^\prime}+k\cos\theta_{\bf k}}{k^\prime\cos\theta_{{\bf k}^\prime}+q\cos\theta^{\prime\prime}_{\bf q}}\right]\ ,\\
\label{boundary3p}
r(k_x,k^\prime_x,q_x\,\vert s,s^\prime,s^{\prime\prime})&=-\left[\frac{s\Gamma_s(k)}{s^\prime\Gamma_{s^\prime}(k^\prime)}\right]\,\texttt{e}^{i(\theta_{\bf k}+\theta_{{\bf k}}^\prime)}\left[\frac{k\cos\theta_{\bf k}-q\cos\theta^{\prime\prime}_{\bf q}}{k^\prime\cos\theta_{{\bf k}^\prime}+q\cos\theta^{\prime\prime}_{\bf q}}\right]\ ,
\end{align}
\end{subequations}
where $r(k_x,k^\prime_x,q_x\,\vert s,s^\prime,s^{\prime\prime})$ depends on $k^\prime_x,\,q_x$ and $s^\prime,\,s^{\prime\prime}$ if $V_0\neq 0$ in addition to its dependence on $s$ and $k_x$ corresponding to $E_K=E_s(k)$. 
In Eqs.\,\eqref{boundary3} and \eqref{boundary3p}, $\theta^{\prime\prime}_{\bf q}$ depends implicitly on the branch index $s^{\prime\prime}$ for given $E_s(k)=E_K$ and $\theta_{\bf k}$. 
It is important to note that the condition $E_s(k)=E_K$ does not determine $k$ uniquely, and therefore, we should assign $k$ and $s$ simultaneously for incidence of electrons instead of $E_K$ solely. 
For $s=s^\prime=s^{\prime\prime}$ and $V_0=0$, we get $q=k^\prime=k$, $\Gamma_{s^{\prime\prime}}(q)=\Gamma_{s^\prime}(k^\prime)=\Gamma_s(k)$ and $\theta_{\bf k}=\theta_{{\bf k}^\prime}=\theta^{\prime\prime}_{\bf q}$ since $q_y=k_y$, which leads to $t=1$ and $r=0$ in Eqs.\,\eqref{boundary3} and \eqref{boundary3p}. On the other hand, even if $V_0\to 0$ but $s\neq s^\prime,\,s^{\prime\prime}$, we still obtain $r\neq 0$ 
due to inter-spin-channel for tunneling and spin-procession for reflection. Here, the spin-procession process becomes possible for reflection and tunneling of electrons because of the spin-state mixing by 
Rashba-Zeeman coupling.
\medskip

\begin{figure} 
\centering
\includegraphics[width=0.65\textwidth]{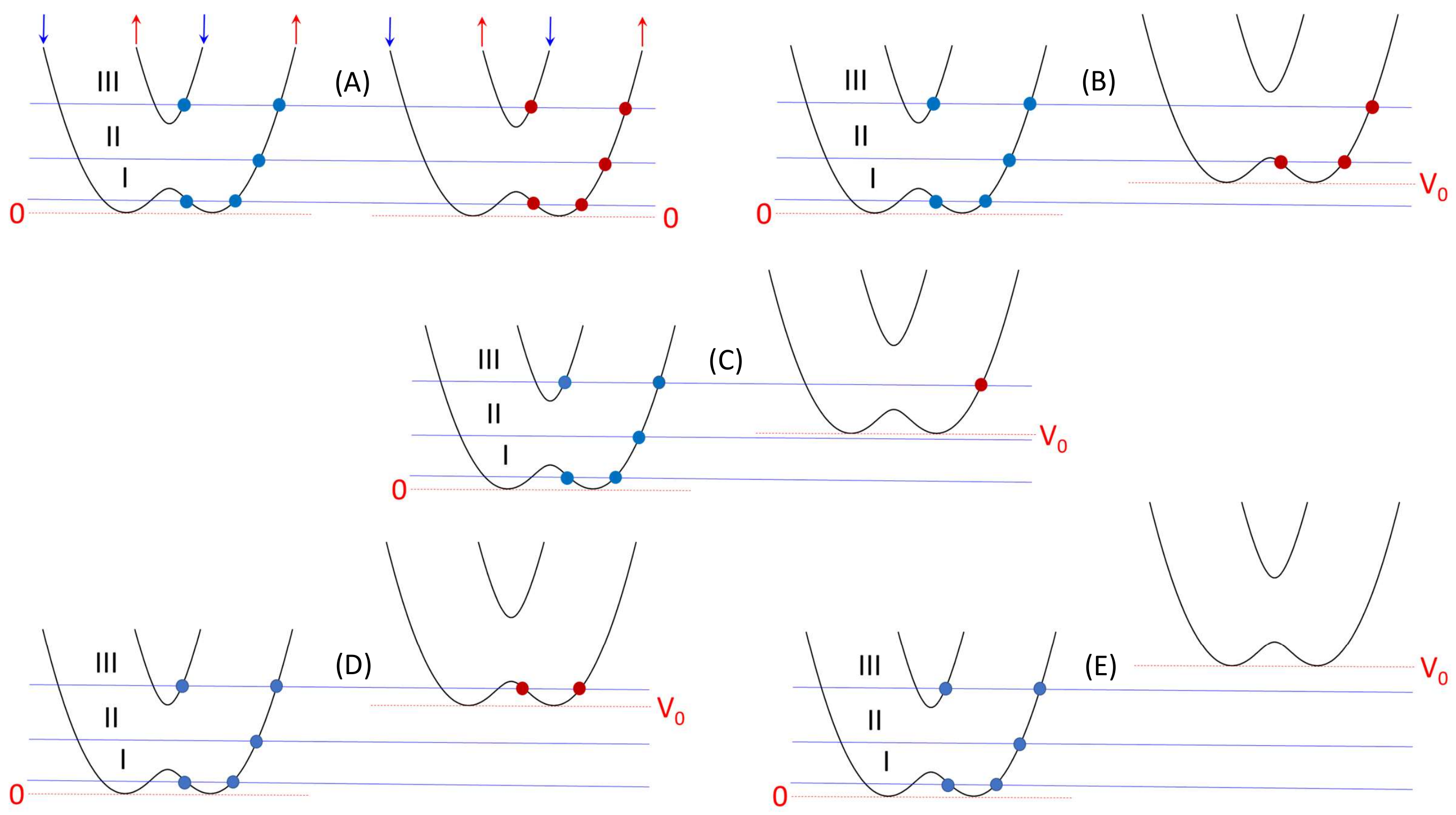}
\caption{Energy dispersions on the incidence (left) and the step-barrier (right) sides with a barrier height $V_0$, where three different incident kinetic energies of $E_K$ (blue horizontal lines) represent energy ranges (I), (II) and (III). The blue and red dots refer to selected electronic states in ranges I, II and III for the incident $(k)$ and barrier $(q)$ sides, respectively.  
(A) $V_0=0$ with aligned $E_K$ states and ranges on the barrier side, where red-upward (blue-downward) arrows indicate up-spin (down-spin) states of electrons; 
(B) $V_0>0$ with allowed $E_K$ states in staggered ranges I, II on the barrier side; 
(C) $V_0>0$ with allowed $E_K$ states in staggered range II on the barrier side; 
(D) $V_0>0$ with allowed $E_K$ states in staggered range I on the barrier side; 
(E) $V_0>E_K$ with no allowed $E_K$ state in all ranges on the barrier side.}
\label{FIG1}
\end{figure}

By using the results in Eqs.\,\eqref{boundary3} and \eqref{boundary3p}, the transmission ${\cal T}(\mbox{\boldmath$k$},\mbox{\boldmath$k$}^\prime,\mbox{\boldmath$q$}\,\vert s,s^\prime,s^{\prime\prime})$ and reflection ${\cal R}(\mbox{\boldmath$k$},\mbox{\boldmath$k$}^\prime,\mbox{\boldmath$q$}\,\vert s,s^\prime,s^{\prime\prime})$ coefficients are obtained as 

\begin{subequations}
\begin{align}
\label{BigT}
{\cal T}(\mbox{\boldmath$k$},\mbox{\boldmath$k$}^\prime,\mbox{\boldmath$q$}\,\vert s,s^\prime,s^{\prime\prime})&= \left|\frac{J_x^{\rm (t)}(s^{\prime\prime},\mbox{\boldmath$q$};\,|t|^2) }{J_x^{\rm (i)} (s,\mbox{\boldmath$k$})}\right|
= |t(k_x,k^\prime_x,q_x\,\vert s,s^\prime,s^{\prime\prime})|^2  \left|\frac{I_{s^{\prime\prime}}(q)\cos\theta^{\prime\prime}_{\bf q}}
{I_s(k)\cos\theta_{\bf k}}  \right|\ ,\\
\label{BigR}
{\cal R}(\mbox{\boldmath$k$},\mbox{\boldmath$k$}^\prime,\mbox{\boldmath$q$}\,\vert s,s^\prime,s^{\prime\prime})&=\left|\frac{J_x^{\rm (r)}(s^\prime,\mbox{\boldmath$k$}^\prime;\,|r|^2)}{J_x^{\rm (i)} (s,\mbox{\boldmath$k$})}\right|=\left|r\right(k_x,k^\prime_x,q_x\,\vert s,s^\prime,s^{\prime\prime})|^2\left|\frac{I_{s^{\prime}}(k^\prime)\cos\theta_{{\bf k}^\prime}}
{I_s(k)\cos\theta_{\bf k}}  \right|\ ,
\end{align}
\end{subequations}
where ${\cal T}(\mbox{\boldmath$k$},\mbox{\boldmath$k$}^\prime,\mbox{\boldmath$q$}\,\vert s,s^\prime,s^{\prime\prime})+{\cal R}(\mbox{\boldmath$k$},\mbox{\boldmath$k$}^\prime,\mbox{\boldmath$q$}\,\vert s,s^\prime,s^{\prime\prime})$ is usually not unity since it only represents the probability current in the $x$ direction. 
In Eqs.\,\eqref{BigT} and \eqref{BigR}, for fixed branch indexes $s,\,s^\prime,\,s^{\prime\prime}$, incident energy $E_K=E_s(k)$ and $k_y$ (or $\theta_{\bf k}$), $k_x$, $q_x$, $k^\prime_x$ for given $k_y$ and $\alpha^{\prime\prime}=\alpha$ are calculated as 

\begin{subequations}
\begin{align}
\label{kq1}
k_\pm&\equiv\sqrt{k_x^2+k_y^2}=\sqrt{2}k_R\left\{1+\frac{E_K+E_{\rm min}}{2E_R}\pm\sqrt{1+\frac{E_K+E_{\rm min}}{E_R}+\frac{\Delta_Z^2}{4E_R^2}}\ \right\}^{1/2}\ ,\\
\label{kq2}
q_\pm&\equiv\sqrt{q_x^2+k_y^2}=\sqrt{2}k_R\left\{1+\frac{E_K-V_0+E_{\rm min}}{2E_R}\pm\sqrt{1+\frac{E_K-V_0+E_{\rm min}}{E_R}+\frac{\Delta_Z^2}{4E_R^2}}\ \right\}^{1/2}\ ,\\
\label{kq3}
k^\prime_\pm&\equiv\sqrt{k_x^{\prime 2}+k_y^2}=\sqrt{2}k_R\left\{1+\frac{E_K+E_{\rm min}}{2E_R}\pm\sqrt{1+\frac{E_K+E_{\rm min}}{E_R}+\frac{\Delta_Z^2}{4E_R^2}}\ \right\}^{1/2}\ ,
\end{align}
\end{subequations}
where $k^\prime_\pm$ can be either same as or different from $k_\pm$ if more than one tunneling or reflection channel exist, 
$k_+$ ($q_+$) and $k_-$ ($q_-$) represent, separately, the larger and smaller $k$ ($q$) values, 
$E_R=\alpha_R^2/4\alpha$ is the rashba energy, $k_R=\alpha_R/2\alpha\hbar$ is the Rashba wave number, and $E_{\rm min}=-E_R(1+\Delta_Z^2/E^2_R)$ for $\overline{E}_s(k)$ at $k_{\rm min}=\pm k_R\sqrt{1-\Delta_Z^2/4E_R^2}$. The results in Eqs.\,\eqref{kq1}$-$\eqref{kq3} are illustrated by Fig.\,\ref{FIG1}.
\medskip

For a given incident energy $E_K$, if multiple intersection points of wave number exist for $E_s(k)=E_K$ on both sides of incidence and step barrier, from the conservation of total particle number, 
we acquire 

\begin{equation}
\sum\limits_{s,s^\prime,s^{\prime\prime}}\,\left[{\cal T}(\mbox{\boldmath$k$},\mbox{\boldmath$k$}^\prime,\mbox{\boldmath$q$}\,\vert s,s^\prime,s^{\prime\prime})+{\cal R}(\mbox{\boldmath$k$},\mbox{\boldmath$k$}^\prime,\mbox{\boldmath$q$}\,\vert s,s^\prime,s^{\prime\prime})\right]=\sum\limits_{s,s^\prime,s^{\prime\prime}}\,1\equiv N_{\rm T}\ ,
\label{add1}
\end{equation}
where $N_{\rm T}$ stands for the total number of inequivalent cases under the condition of $E_s(k)=E_K$ for different incidence, reflection and tunneling channels. Then, by defining average transmission 
coefficient ${\cal T}_{\rm av}(\mbox{\boldmath$k$},\mbox{\boldmath$k$}^\prime,\mbox{\boldmath$q$})$ and reflection coefficient ${\cal R}_{\rm av}(\mbox{\boldmath$k$},\mbox{\boldmath$k$}^\prime,\mbox{\boldmath$q$})$ 
coefficients through

\begin{eqnarray}
\label{add2}
&&{\cal T}_{\rm av}(\mbox{\boldmath$k$},\mbox{\boldmath$k$}^\prime,\mbox{\boldmath$q$})=\frac{1}{N_{\rm T}}\sum\limits_{s,s^\prime,s^{\prime\prime}}\,{\cal T}(\mbox{\boldmath$k$},\mbox{\boldmath$k$}^\prime,\mbox{\boldmath$q$}\,\vert s,s^\prime,s^{\prime\prime})\ ,\\
\label{add3}
&&{\cal R}_{\rm av}(\mbox{\boldmath$k$},\mbox{\boldmath$k$}^\prime,\mbox{\boldmath$q$})=\frac{1}{N_{\rm T}}\sum\limits_{s,s^\prime,s^{\prime\prime}}\,{\cal R}(\mbox{\boldmath$k$},\mbox{\boldmath$k$}^\prime,\mbox{\boldmath$q$}\,\vert s,s^\prime,s^{\prime\prime})\ ,\ 
\end{eqnarray}
we arrive from Eq.\,\eqref{add1} at the conclusion that

\begin{equation}
{\cal T}_{\rm av}(\mbox{\boldmath$k$},\mbox{\boldmath$k$}^\prime,\mbox{\boldmath$q$})+{\cal R}_{\rm av}(\mbox{\boldmath$k$},\mbox{\boldmath$k$}^\prime,\mbox{\boldmath$q$})\equiv 1\ .
\label{add4}
\end{equation}
Consequently, we know from Eq.\,\eqref{add4} that ${\cal T}_{\rm av}(\mbox{\boldmath$k$},\mbox{\boldmath$k$}^\prime,\mbox{\boldmath$q$})\leq 1$ and 
${\cal R}_{\rm av}(\mbox{\boldmath$k$},\mbox{\boldmath$k$}^\prime,\mbox{\boldmath$q$})\leq 1$ must be satisfied physically. 
Moreover, we refer ${\cal T}(\mbox{\boldmath$k$},\mbox{\boldmath$k$}^\prime,\mbox{\boldmath$q$}\,\vert s,s^\prime,s^{\prime\prime})$ in Eq.\,\eqref{add2} and ${\cal R}(\mbox{\boldmath$k$},\mbox{\boldmath$k$}^\prime,\mbox{\boldmath$q$}\,\vert s,s^\prime,s^{\prime\prime})$ in Eq.\,\eqref{add3} as the partial transmission and reflection coefficients, respectively, 
which can be larger than one as long as ${\cal T}(\mbox{\boldmath$k$},\mbox{\boldmath$k$}^\prime,\mbox{\boldmath$q$}\,\vert s,s^\prime,s^{\prime\prime})/N_{\rm T}$ and ${\cal R}(\mbox{\boldmath$k$},\mbox{\boldmath$k$}^\prime,\mbox{\boldmath$q$}\,\vert s,s^\prime,s^{\prime\prime})/N_{\rm T}$ do not exceed unity.
\medskip
 
For Fig.\,\ref{FIG1}, as $V_0>E_K+E_{\rm min}+2E_R+\Delta_Z^2/2E_R$, $q_\pm$ becomes an imaginary number, implying no tunneling channel is available in the system. 
If $V_0=0$ and $s=s^\prime=s^{\prime\prime}$ for an intra-spin-channel tunneling, we have $q_x=k_x=k_x^\prime$. In this case, we find from Eqs.\,\eqref{boundary3}, \eqref{boundary3p}, \eqref{BigT} and \eqref{BigR} that 
$t\to 1$, $r\to 0$, ${\cal T}(\mbox{\boldmath$k$},\mbox{\boldmath$k$}^\prime,\mbox{\boldmath$q$}\,\vert s,s^\prime,s^{\prime\prime})\to 1$ and ${\cal R}(\mbox{\boldmath$k$},\mbox{\boldmath$k$}^\prime,\mbox{\boldmath$q$}\,\vert s,s^\prime,s^{\prime\prime})\to 0$ simultaneously. If $V_0=0$ but $s\neq s^{\prime\prime}$ and $s\neq s^\prime$ for forward 
and backward spin processions, on the other hand, we get $q_x\neq k_x$ and $k_x^\prime\neq k_x$.
\medskip

As a special gapless case with $\Delta_Z=0$, we get directly from Eqs.\,\eqref{kq1}, \eqref{kq2} and \eqref{kq3} that 

\begin{subequations}
\begin{align}
\label{kq4}
k^{(0)}_\pm&\equiv\sqrt{k_x^2+k_y^2}=\sqrt{2}k_R\left\{1+\frac{E_K+E_{\rm min}}{2E_R}\pm\sqrt{1+\frac{E_K+E_{\rm min}}{E_R}}\ \right\}^{1/2}\ ,\\
\label{kq5}
q^{(0)}_\pm&\sqrt{q_x^2+k_y^2}=\sqrt{2}k_R\left\{1+\frac{E_K-V_0+E_{\rm min}}{2E_R}\pm\sqrt{1+\frac{E_K-V_0+E_{\rm min}}{E_R}}\ \right\}^{1/2}\ ,\\
\label{kq6}
k^{\prime(0)}_\pm&\equiv\sqrt{k_x^{\prime 2}+k_y^2}=\sqrt{2}k_R\left\{1+\frac{E_K+E_{\rm min}}{2E_R}\pm\sqrt{1+\frac{E_K+E_{\rm min}}{E_R}}\ \right\}^{1/2}\ ,\\
\end{align}
\end{subequations}
which always present two nonzero intersection points for wave numbers $k$, $k^\prime$ and $q$ except for a degeneracy case with $E_K=-E_{\rm min}=E_R$ or $E_K=V_0-E_{\rm min}=V_0+E_R$. 
Explicitly, from Eqs.\,\eqref{BigT} and \eqref{BigR} we get

\begin{eqnarray}
\label{new1}
{\cal R}(\mbox{\boldmath$k$},\mbox{\boldmath$k$}^\prime,\mbox{\boldmath$q$}\,\vert s,s^\prime,s^{\prime\prime})&=&\left[\frac{\Gamma_s(k)}{\Gamma_{s^{\prime}}(k^\prime)}\right]^2\,\left[\frac{k\cos\theta_{\bf k}-q\cos\theta^{\prime\prime}_{\bf q}}{k^\prime\cos\theta_{{\bf k}^\prime}+q\cos\theta^{\prime\prime}_{\bf q}}\right]^2\left|\frac{[\alpha\hbar k^\prime+s^{\prime}\Gamma_{s^{\prime}}(k^\prime)\alpha_R]\cos\theta_{{\bf k}^\prime}}{[\alpha\hbar k+s\Gamma_{s}(k)\alpha_R]\cos\theta_{\bf k}}\right|\ ,\\
{\cal T}(\mbox{\boldmath$k$},\mbox{\boldmath$k$}^\prime,\mbox{\boldmath$q$}\,\vert s,s^\prime,s^{\prime\prime})&=&\left[\frac{\Gamma_s(k)}{\Gamma_{s^{\prime\prime}}(q)}\right]^2\,\left[\frac{k^\prime\cos\theta_{{\bf k}^\prime}+k\cos\theta_{\bf k}}{k^\prime\cos\theta_{{\bf k}^\prime}+q\cos\theta^{\prime\prime}_{\bf q}}\right]^2\left|\frac{[\alpha\hbar q+s^{\prime\prime}\Gamma_{s^{\prime\prime}}(q)\alpha_R]\cos\theta^{\prime\prime}_{\bf q}}{[\alpha\hbar k+s\Gamma_{s}(k)\alpha_R]\cos\theta_{\bf k}}\right|\ ,
\label{new2}
\end{eqnarray}
where $\Gamma_s(k)=\gamma_s(k)/[1+\gamma_s^2(k)]$ and $\gamma_s(k)=\alpha_R\hbar k/[\sqrt{\alpha^2_R\hbar^2k^2+\Delta^2_Z}+s\Delta_Z]$.
\medskip

Next, we further include a $\delta$-function potential with an amplitude $d_0$ (in unit of $\hbar\alpha_R=2k_R\hbar^2\alpha$) at the interface and write down $V_B(x)=V_0\Theta_0(x)+d_0\delta(x)$. This leads to a  
discontinuity in the derivative of a wave function. As a result, the boundary conditions in Eqs.\,\eqref{boundary2} and \eqref{boundary2p} for $\alpha=\alpha^{\prime\prime}$ are generalized to

\begin{subequations}
\begin{align}
\label{boundary4}
-is^{\prime\prime}\Gamma_{s^{\prime\prime}}(q)\,\texttt{e}^{i\theta^{\prime\prime}_{\bf q}}&=-is\Gamma_s(k)\,\texttt{e}^{i\theta_{\bf k}}+is^\prime\Gamma_{s^\prime}(k^\prime)\,\texttt{e}^{-i\theta_{{\bf k}^\prime}}\,r\ ,\\
\label{boundary4p}
q_xs^{\prime\prime}\Gamma_{s^{\prime\prime}}(q)\,\texttt{e}^{i\theta_{\bf q}^{\prime\prime}}\,t&=k_xs\Gamma_s(k)\,\texttt{e}^{i\theta_{\bf k}}+k^\prime_xs^\prime\Gamma_{s^\prime}(k^\prime)\,\texttt{e}^{-i\theta_{{\bf k}^\prime}}\,r
+\left(\frac{d_0}{\hbar^2\alpha}\right)is^{\prime\prime}\Gamma_{s^{\prime\prime}}(q)\,\texttt{e}^{i\theta^{\prime\prime}_{\bf q}}\,t\ ,
\end{align}
\end{subequations}
This produces the solutions that

\begin{subequations}
\begin{align}
\label{boundary5}
t(k_x,k^\prime_x,q_x\,\vert s,s^\prime,s^{\prime\prime};d_0)&=\left[\frac{s\Gamma_s(k)}{s^{\prime\prime}\Gamma_{s^{\prime\prime}}(q)}\right]\,\texttt{e}^{i(\theta_{\bf k}-\theta^{\prime\prime}_{\bf q})}\,\left[\frac{k^\prime\cos\theta_{{\bf k}^\prime}+k\cos\theta_{\bf k}}{k^\prime\cos\theta_{{\bf k}^\prime}+q\cos\theta^{\prime\prime}_{\bf q}-i(d_0/\hbar^2\alpha)}\right]\ ,\\
\label{boundary5p}
r(k_x,k^\prime_x,q_x\,\vert s,s^\prime,s^{\prime\prime};d_0)&=-\left[\frac{s\Gamma_s(k)}{s^{\prime}\Gamma_{s^{\prime}}(k^\prime)}\right]\,\texttt{e}^{i(\theta_{\bf k}+\theta_{{\bf k}^\prime})}\,\left[\frac{k\cos\theta_{\bf k}-q\cos\theta^{\prime\prime}_{\bf q}+i(d_0/\hbar^2\alpha)}{k^\prime\cos\theta_{{\bf k}^\prime}+q\cos\theta^{\prime\prime}_{\bf q}-i(d_0/\hbar^2\alpha)}\right]\ .
\end{align}
\end{subequations}
In this case, however, we find $r\neq 0$ and $|t|<1$ even for $V_0=0$, $q=k^\prime=k$ and $\theta_{\bf k}=\theta_{{\bf k}^\prime}=\theta^{\prime\prime}_{\bf q}$ due to the presence of a $\delta$-function potential ($d_0\neq 0$).
\medskip

From the physics perspective, for non-magnetic materials, the electrical control of inter-spin-channel tunneling by Rashba interaction demonstrates a very interesting and unique mechanism and becomes ideal for applications in spintronics. It is believed that such a phenomenon depends sensitively on the selections of branch indexes $s$ and $s^{\prime\prime}$, incident kinetic energy $E_K$, incident angle $\theta_{\bf k}$, inverse effective mass $\alpha$, Zeeman gap $\Delta_Z$, barrier height $V_0$, Rashba spin-orbit parameter $\alpha_R$ and even the amplitude $d_0$ of a $\delta$-function potential.
\medskip

More interestingly, for a specifically selected incident kinetic energy $E_K$ of an electron and its corresponding $\{s,k\}$ electronic state, two relevant $\{s^{\prime\prime},\mbox{\boldmath$q$}_{1,2}\}$ states of the lower branch $s^{\prime\prime}=-1$ (or the lower and upper branches $s^{\prime\prime}=\pm 1$) exist within the barrier region as $V_0\neq 0$, e.g., $s=-s^{\prime\prime}$ in Fig.\,\ref{FIG1}D. 
Therefore, for a given incident angle $\theta_{\bf k}$ or $k_y$ 
and a fixed $k$ by $E_K$ as well, two equivalent electron tunneling processes can coexist simultaneously in the barrier region with different diffraction angles $\theta''_{{\bf q}_1}$ and $\theta''_{{\bf q}_2}$. Consequently, these diffracted electron beams with different exit angles are expected to interference with each other, similar to a double-slit experiment for a normally-incident light wave. Here, the spatial distribution of a total probability density $\left|\Psi_{\rm tot}^{\rm (t)}(\mbox{\boldmath$r$})\right|^2$ of electrons on the step-barrier side takes the form 

\begin{eqnarray}
\nonumber
{\cal A}\left|\Psi_{\rm tot}^{\rm (t)}(\mbox{\boldmath$r$})\right|^2&=&\frac{1}{2N_T}\,\left|t_1(k_x,k^\prime_x,q_{1x}\,\vert s,s^\prime,s^{''}_1)\,\frac{\texttt{e}^{iq_{1x}x}}{\sqrt{1+\gamma_{s_1''}^2(q_1)}}
\left[\begin{array}{c}
-is_1''\gamma_{s_1''}(q_1)\texttt{e}^{i\theta^{\prime\prime}_{{\bf q}_1}}\\
\\
1
\end{array}\right]\right.\\
\label{new-11}
&+&\left.t_2(k_x,k^\prime_x,q_{2x}\,\vert s,s^\prime,s^{''}_2)\,\frac{\texttt{e}^{iq_{2x}x}}{\sqrt{1+\gamma_{s_2''}^2(q_2)}}
\left[\begin{array}{c}
-is_2''\gamma_{s_2''}(q_2)\texttt{e}^{i\theta^{\prime\prime}_{{\bf q}_2}}\\
\\
1
\end{array}\right]\right|^2\ ,
\end{eqnarray}
which includes the unique spin-state quantum interference between two correlated coherent electron waves. Similar conclusion can be drawn for the total reflection probability density $\left|\Psi_{\rm tot}^{\rm (r)}(\mbox{\boldmath$r$})\right|^2$ of electrons outside the step region, written as

\begin{eqnarray}
\nonumber
{\cal A}\left|\Psi_{\rm tot}^{\rm (r)}(\mbox{\boldmath$r$})\right|^2&=&\frac{1}{2N_T}\,\left|r_1(k_x,k^\prime_{1x},q_{x}\,\vert s,s_1^\prime,s^{''})\,\frac{\texttt{e}^{-ik^\prime_{1x}x}}{\sqrt{1+\gamma_{s_1^\prime}^2(k^\prime_1)}}
\left[\begin{array}{c}
-is_1^\prime\gamma_{s_1^\prime}(k^\prime_1)\texttt{e}^{i(\pi-\theta_{{\bf k}^\prime_1})}\\
\\
1
\end{array}\right]\right.\\
\label{new-12}
&+&\left.r_2(k_x,k^\prime_{2x},q_{x}\,\vert s,s_2^\prime,s^{''})\,\frac{\texttt{e}^{-ik^\prime_{2x}x}}{\sqrt{1+\gamma_{s_2'}^2(k^\prime_2)]}}
\left[\begin{array}{c}
-is_2^\prime\gamma_{s_2^\prime}(k^\prime_2)\texttt{e}^{i(\pi-\theta_{{\bf k}^\prime_2})}\\
\\
1
\end{array}\right]\right|^2\ .
\end{eqnarray}
It is interesting to note from Eqs.\,\eqref{new-11} and \eqref{new-12} that the degree of such spin-state quantum interference can be controlled electrically by their tunneling amplitudes $t_{1,2}$ and diffraction angles $\theta^{\prime\prime}_{{\bf q}_{1,2}}$ or reflection amplitudes $r_{1,2}$ and angles $\theta_{{\bf k}^\prime_{1,2}}$, respectively. Such a unique property is expected acquiring a lot of implications for spintronics and quantum-computation devices.

\section{Numerical Results and Discussion} 
\label{sec4}

\begin{table}[htbp]
\caption{\bf Parameters used for calculations}
\begin{tabular}{cccc}
\hline\hline
Parmeter\,\cite{param1,param2} 				&	Description	& 	Value	& 	Units \\
\hline
$k_R^c$           &   Rashba wave number              &   0.015                   &   \AA$^{-1}$          \\              
$k_R^v$           &   Rashba wave number              &   0.015                   &   \AA$^{-1}$          \\              
%$\hbar\alpha_R^c$				&	Rashba parameter						&	16.0					&	eV$\cdot$\AA				\\
%$\hbar\alpha_R^v$				&	Rashba parameter						&	30.7					&	eV$\cdot$\AA				\\
%$m_c^\ast$					&	effective mass 		&\ \ \ \	0.28\ \ \ \ \  	&  9.1$\times$10$^{-31}\,$kg					\\
%$|m_v^\ast |$					&	effective mass 		&\ \ \ \	0.15\ \ \ \ \  	&  9.1$\times$10$^{-31}\,$kg					\\
$E^c_R$           &   Rashba energy                   &   118                     &   meV                 \\
$E^v_R$           &   Rashba energy                   &   235                     &   meV                 \\
$2\Delta_Z$            &   bandgap                      &\ \ \ \   $50$\,\cite{param2}          &              meV     \\
$d_0$     & amplitude   & 2    &   $2E_R^c/k^c_R$\\
\hline\hline
\end{tabular}
\label{tab-1}
\end{table}

The parameters used in numerical computations are listed in Table\ \ref{tab-1}. Other parameters, such as $V_0$, $E_K$, $\theta_{\bf k}$ and $d_0$, will be given directly in figures. 
In Sec.\ \ref{sec4}, we start with discussing various cases without a step-potential barrier (i.e. $V_0\to 0$) in Sec.\ \ref{sec-4.1}, from which we demonstrate the spin-procession dynamics for both forward and backward motions of a spin-polarized electron due to Rashba-Zeeman coupling and spin-mixing effects in the system. In Sec.\ \ref{sec-4.2}, we further study the dynamics for tunneling and reflection of a spin-polarized incident electron under a step-potential barrier in the system, which exhibits spin-state interference in either reflected or transmitted 
pair of mixed-spin-state electrons with different reflection or diffraction angles.
\medskip

\subsection{Spin-Procession Dynamics For $V_0\to 0$}
\label{sec-4.1}

For numerical computations, we first divide the incident energy range of $E_K$ into three sub-ranges, (I), (II) and (III) for energy dispersion $E_s(k)$, as illustrated in Fig.\,\ref{FIG1}. 
In range I, we have $0\leq E_K\leq |E_{\rm min}|-\Delta_Z$ below the lower edge of the Zeeman gap, where $|E_{\rm min}|=E_R(1+\Delta_Z^2/E_R^2)$. 
In range II, on the other hand, we have $|E_{\rm min}|-\Delta_Z\leq E_K\leq |E_{\rm min}|+\Delta_Z$ within the Zeeman gap. 
Finally, in range III, we have $E_K\geq |E_{\rm min}|+\Delta_Z$ above the upper edge of the Zeeman gap. 
Next, for each range of $E_K$, we select different values for the step-barrier height $V_0$ within the interval of $0\leq V_0\leq E_K$. 
Moreover, for given values of $E_K$ and $V_0$, we choose different incident angles within $-\pi/2\leq\theta_{\bf k}\leq\pi/2$, corresponding to various values of $k_y$. 
Furthermore, for fixed $E_K$ and $V_0$ values, there always exist either one or two wave-number intersection points with two Rashba-Zeeman split energy bands in the incident and barrier regions, respectively. This gives rise to values of $q_\pm$ for tunneling in the barrier side while 
$k_\pm$ and $k^\prime_\pm$ for incidence and reflection in the incident side, as calculated by Eqs.\,\eqref{kq1},\,\eqref{kq2} and \eqref{kq3}.
\medskip

\begin{table}[htbp]
\caption{\bf Configurations for $E_K=50\,$meV within range-I and $V_0=0$}
\begin{tabular}{ccccc}
\hline\hline
Case Number\ \ \ &  Incidence\ \ \ & Reflection\ \ \ & Transmission\ \ \ & Tunneling Status \\
\hline
$1$&  $a$&  $a$&  $a^{\prime}$&  $\checkmark$  \\
$2$&  $a$&  $a$&  $b^{\prime}$&  $\checkmark$  \\
$3$&  $a$&  $b$&  $a^{\prime}$&  $\checkmark$  \\
$4$&  $b$&  $a$&  $a^{\prime}$&  $\times$  \\
$5$&  $a$&  $b$&  $b^{\prime}$&  $\times$  \\
$6$&  $b$&  $a$&  $b^{\prime}$&  $\checkmark$  \\
$7$&  $b$&  $b$&  $a^{\prime}$&  $\checkmark$  \\
$8$&  $b$&  $b$&  $b^{\prime}$&  $\checkmark$  \\
\hline\hline
\end{tabular}
\label{tab-2}
\end{table}
\medskip

\begin{figure} 
\centering
\includegraphics[width=0.65\textwidth]{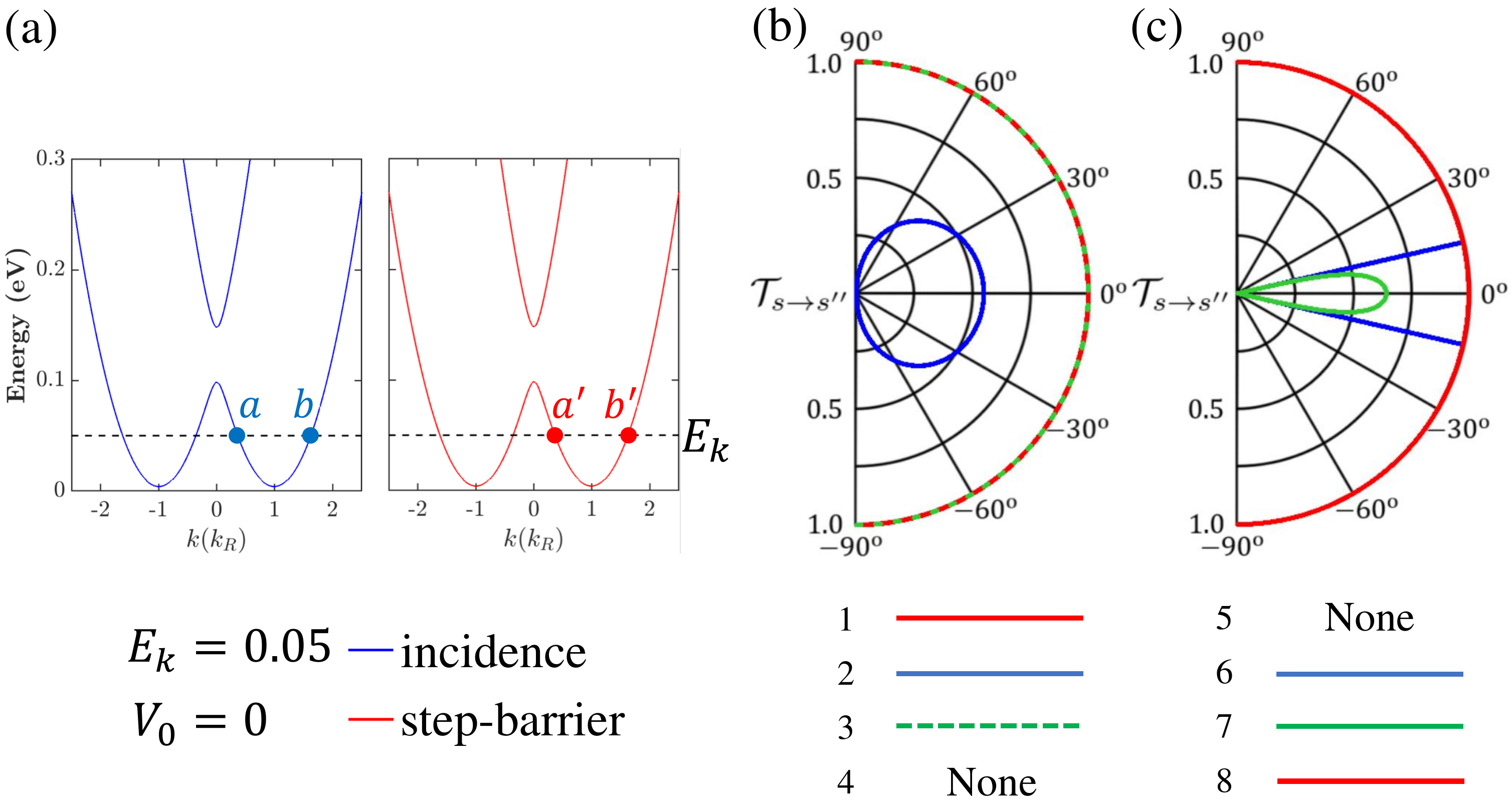}
\caption{(a) Calculated split Rashiba-Zeeman energy bands as a function of scaled electron wave number k/$k_R$, where the incident energy $E_K$ is indicated by a black-dashed line. Symbols $a$, $a^{\prime}$, $b$, $b^{\prime}$ denote four wave-number intersection points for incident energy $E_K$ = 50 meV of electrons within range-I and $V_0$ = 0 or in situation-A of Fig. 1. (b-c) Polar plots of partial tunneling coefficients ${\mathcal{T}}_{s \rightarrow s^{\prime\prime}}$ as a function of incident angle $\theta_k$ for fixed $E_K$ = 50 meV and $V_0$ = 0 in N$_T$ = 8 different cases listed in Table\ \ref{tab-2}.}
\label{FIG2}
\end{figure}

Our numerically computed results are displayed in Fig.\,\ref{FIG2} for all $N_{\rm T}=8$ cases listed in Table\ \ref{tab-2}.	
As seen in Table\ \ref{tab-2}, for Case-4 we have $k=k_+$, $k^\prime=q=k_-\ll k$, and therefore $\theta_{\bf k}\ll\theta_{{\bf k}'}=\theta^{\prime\prime}_{\bf q}$ since $k_y=k_y'=q_y$. 
As a result, we know from Eq.\,\eqref{boundary3} that $|t(k_x,k^\prime_x,q_x\,\vert s,s^\prime,s^{\prime\prime})|/\sqrt{N_{\rm T}}\gg 1$ in Case-4, which implies that such a forward partial 
spin procession should be excluded. 
On the other hand, for Case-5 we have $k=k_-$, $k^\prime=q=k_+\gg k$, leading to $\theta_{\bf k}\gg\theta_{{\bf k}'}=\theta^{\prime\prime}_{\bf q}$ due to $k_y=k_y'=q_y$. 
Consequently, we find ${\cal T}_{s\to s^{\prime\prime}}/N_{\rm T}\gg 1$ from Eq.\,\eqref{BigT}, which indicates that such a forward partial spin procession should also be ruled out. 
\medskip

For Case-1 in Fig.\,\ref{FIG2}, we have a simple situation with $k=k^\prime=q$ or $\theta_{\bf k}=\theta_{{\bf k}^\prime}=\theta^{\prime\prime}_{\bf q}$. In this case, it is clear from Eqs.\,\eqref{boundary3} and \eqref{BigT} that ${\cal T}_{s\to s^{\prime\prime}}\equiv 1$ but with no spin procession involved. 
Similar conclusion can be drawn for Cases-3 \& 8 without having a spin procession. For Case-6, although we have ${\cal T}_{s\to s^{\prime\prime}}\equiv 1$ with no spin procession, it is restricted by $\theta_{\bf k}<\sin^{-1}(k_-/k_+)\equiv\theta_{\rm c}$ 
due to the constraint $k_x^{\prime 2}=k_-^2-k_y^2=k_-^2-k_+^2\sin^2\theta_{\bf k}\geq 0$ for the incident angle $\theta_{\bf k}$, and then, no forward motion of electrons is expected 
in this case if $\theta_{\bf k}>\theta_{\rm c}$.
Finally, for Case-2 we acquire $k=k^\prime=k_-$ but $q=k_+\gg k$. This leads to $\theta_{\bf k}=\theta_{{\bf k}^\prime}\gg\theta^{\prime\prime}_{\bf q}$. Consequently, 
from Eq.\,\eqref{boundary3} we arrive at a conclusion that $|t(k_x,k^\prime_x,q_x\,\vert s,s^\prime,s^{\prime\prime})|<1$, or a forward partial spin procession should occur without  
a critical angle $\theta_{\rm c}$ in this case.
In a similar Case-7, we get $k=k^\prime=k_+$ while $q=k_-\ll k$, and in this case we still have ${\cal T}_{s\to s^{\prime\prime}}<1$ from Eq.\,\eqref{BigT} 
for a forward partial spin procession but with an angle restriction due to $\theta^{\prime\prime}_{\bf q}\gg\theta_{\bf k}=\theta_{{\bf k}^\prime}$.
\medskip

\begin{figure} 
\centering
\includegraphics[width=0.55\textwidth]{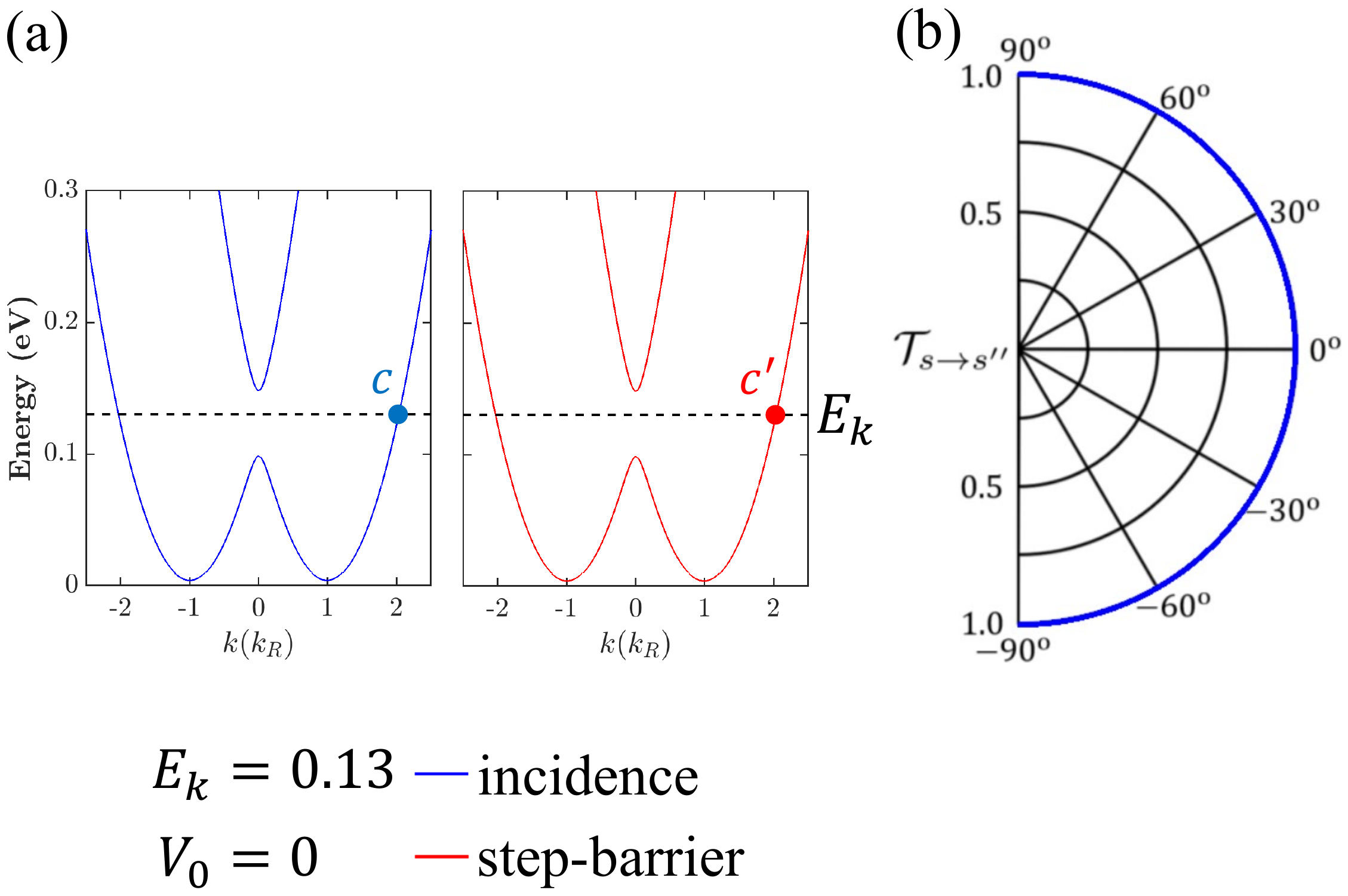}
\caption{(a) Calculated split Rashiba-Zeeman energy bands as a function of scaled electron wave number k/$k_R$, where the incident energy $E_K$ is indicated by a black-dashed line. Symbols $c$, $c^{\prime}$ denote two wave-number intersection points for incident energy $E_K$ = 130 meV of electrons within range-II and $V_0$ = 0 or in situation-A of Fig. 1. (b) Polar plots of partial tunneling coefficients ${\mathcal{T}}_{s \rightarrow s^{\prime\prime}}$ as a function of incident angle $\theta_k$ for fixed $E_K$ = 130 meV and $V_0$ = 0 in a single (N$_T$ = 1) case.}
\label{FIG3}
\end{figure}

Figure\ \ref{FIG3} displays numerical result for a single case with $E_K$ fallen into an energy gap within range II.	
Here, with $k=k^\prime=q=k_+$, we always acquire ${\cal T}_{s\to s^{\prime\prime}}=1$ for all incident angles $\theta_{\bf k}$. 
Clearly, no spin procession is involved in this case for both forward and backward motions of an incident electron. 
\medskip

\begin{table}[htbp]
\caption{\bf Configurations for $E_K=200\,$meV in range-III and $V_0=0$}
\begin{tabular}{ccccc}
\hline\hline
Case Number\ \ \ &  Incidence\ \ \ & Reflection\ \ \ & Transmission\ \ \ & Tunneling Status \\
\hline
$1$&  $d$&  $d$&  $d^{\prime}$&  $\checkmark$  \\
$2$&  $d$&  $d$&  $e^{\prime}$&  $\checkmark$  \\
$3$&  $d$&  $e$&  $d^{\prime}$&  $\checkmark$  \\
$4$&  $e$&  $d$&  $d^{\prime}$&  $\times$  \\
$5$&  $d$&  $e$&  $e^{\prime}$&  $\times$  \\
$6$&  $e$&  $d$&  $e^{\prime}$&  $\checkmark$  \\
$7$&  $e$&  $e$&  $d^{\prime}$&  $\checkmark$  \\
$8$&  $e$&  $e$&  $e^{\prime}$&  $\checkmark$  \\
\hline\hline
\end{tabular}
\label{tab-3}
\end{table}
\medskip

\begin{figure} 
\centering
\includegraphics[width=0.65\textwidth]{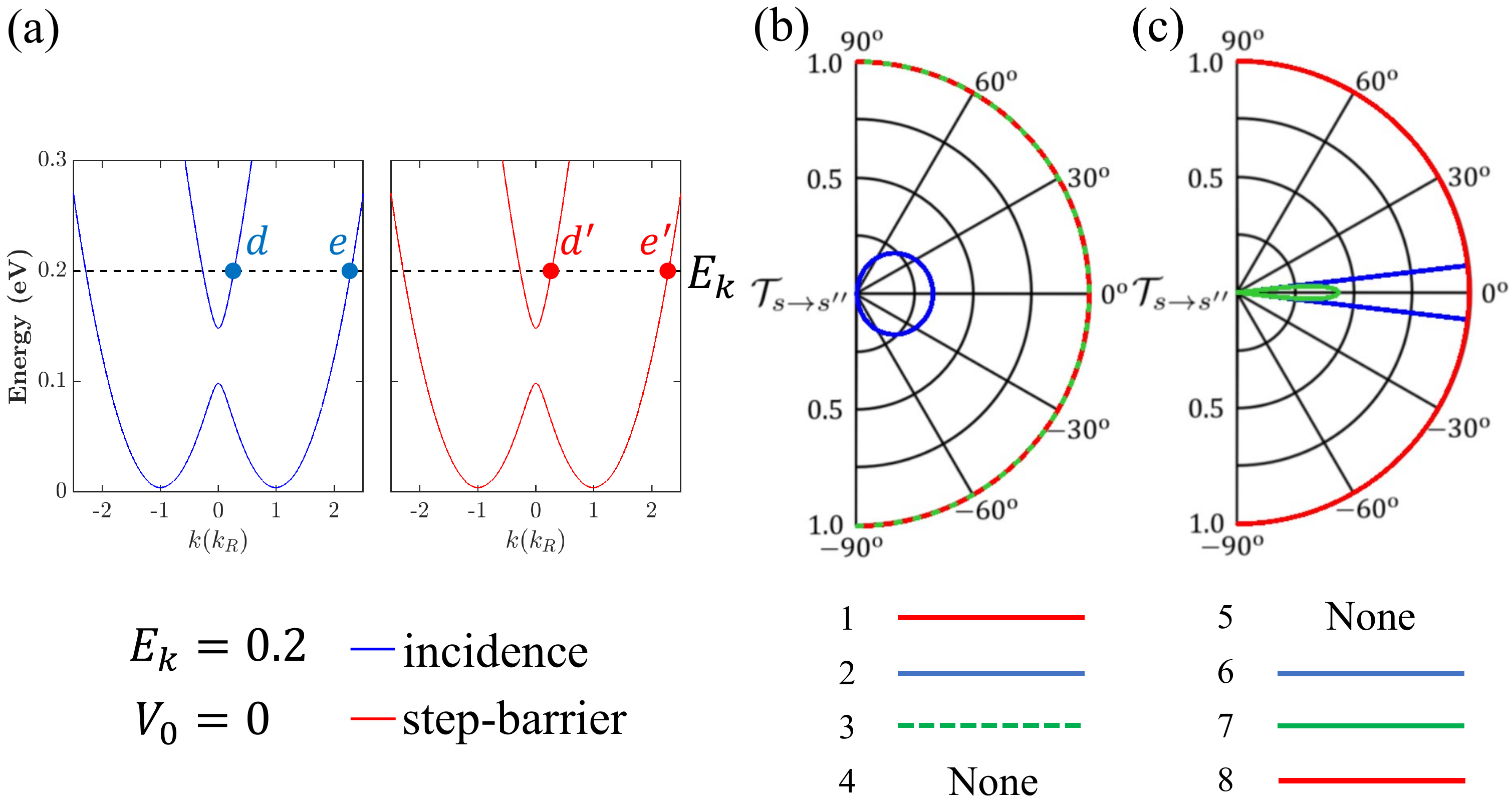}
\caption{(a) Calculated split Rashiba-Zeeman energy bands as a function of scaled electron wave number k/$k_R$, where the incident energy $E_K$ is indicated by a black-dashed line. Symbols $d$, $d^{\prime}$, $e$, $e^{\prime}$ denote four wave-number intersection points for incident energy $E_K$ = 200 meV of electrons within range-III and $V_0$ = 0 or in situation-A of Fig. 1. (b-c) Polar plots of partial tunneling coefficients ${\mathcal{T}}_{s \rightarrow s^{\prime\prime}}$ as a function of incident angle $\theta_k$ for fixed $E_K$ = 200 meV and $V_0$ = 0 in N$_T$ = 8 different cases listed in Table\ \ref{tab-3}.}
\label{FIG4}
\end{figure}

Compared with the results in Fig.\,\ref{FIG2}, for an incident energy $E_K$ sitting within range III in Fig.\,\ref{FIG4} as well as for all $N_{\rm T}=8$ cases listed in Table\ \ref{tab-3}, we find 
that quite similar analysis could be performed. From Fig.\,\ref{FIG4} and Table\ \ref{tab-3}, 
we further know that Cases-3 \& 6 are associated with no spin procession while Cases-2 \& 7 relate to a forward partial spin procession, 
quite similar to those displayed in Fig.\,\ref{FIG2}.
\medskip

Discussions on Figs.\,\ref{FIG2}-\ref{FIG4} are only limited to a forward partial spin-procession process. For a backward partial spin-procession process, similar analysis can also be performed in the same way. 

\subsection{Tunneling-Dynamics For $V_0>0$}
\label{sec-4.2}

In previous Sec.\ \ref{sec-4.1}, we have explored both forward and backward spin-procession physics for a spin-polarized incident electron in the absence of a potential barrier ($V_0\to 0$).
Now, we would consider dynamics for intra-spin-channel (without a spin procession) and inter-spin-channel (with a spin procession) tunnelings of a spin-polarized incident electron in the presence of a step-potential barrier ($V_0>0$) in the system. 
\medskip

\begin{table}[htbp]
\caption{\bf Configurations for $E_K=130\,$meV in range-II and $V_0=50\,$meV}
\begin{tabular}{ccccc}
\hline\hline
Case Number\ \ \ &  Incidence\ \ \ & Reflection\ \ \ & Transmission\ \ \ & Tunneling Status \\
\hline
$1$&  $c$&  $c$&  $a^{\prime}$&  $\checkmark$  \\
$2$&  $c$&  $c$&  $b^{\prime}$&  $\checkmark$  \\
\hline\hline
\end{tabular}
\label{tab-4}
\end{table}
\medskip

\begin{figure} 
\centering
\includegraphics[width=0.55\textwidth]{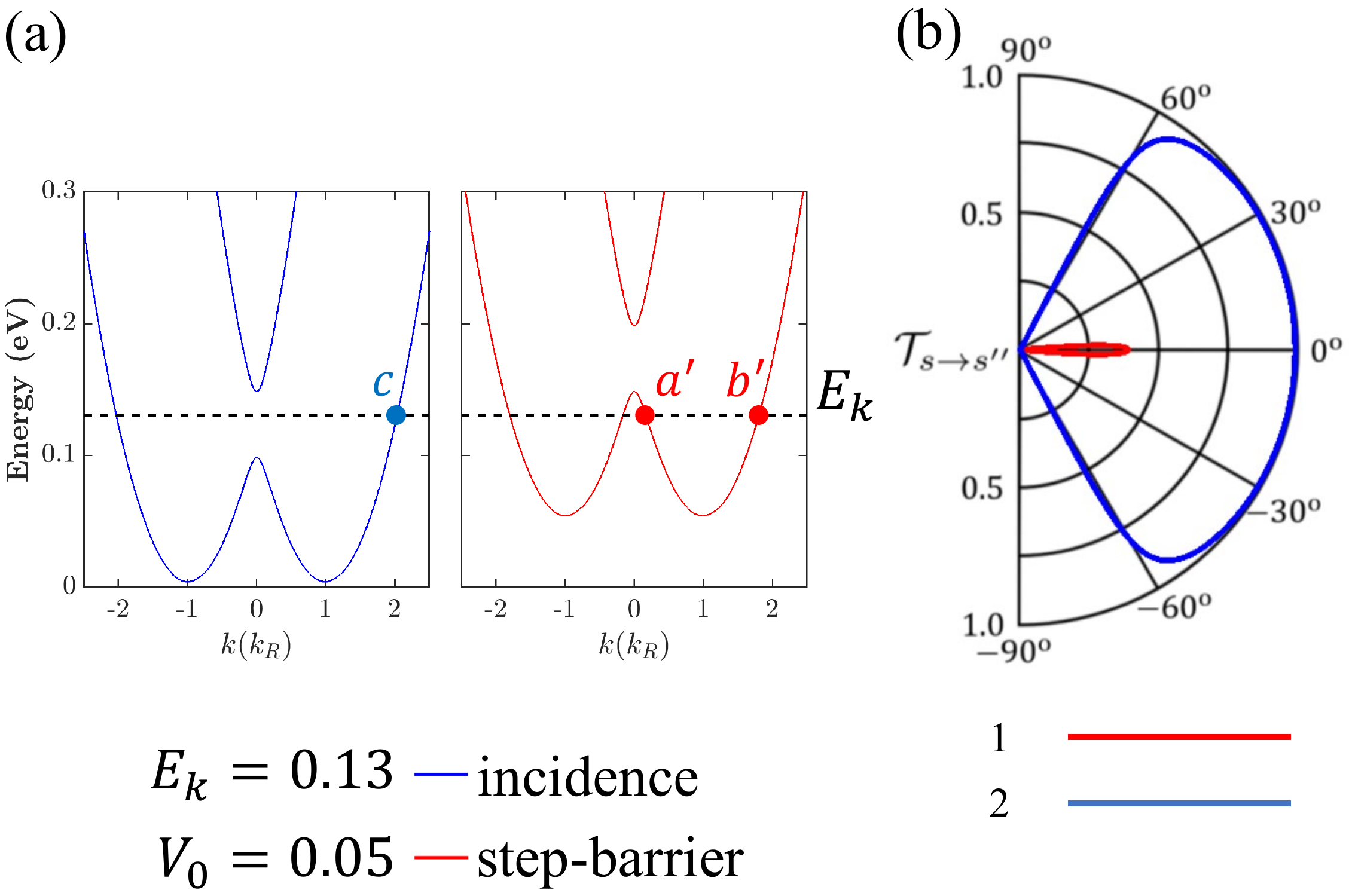}
\caption{(a) Calculated split Rashiba-Zeeman energy bands as a function of scaled electron wave number k/$k_R$, where the incident energy $E_K$ is indicated by a black-dashed line while dispersion curves on the incident and barrier sides are represented by blue and red curves. Symbols $c$, $a^{\prime}$, $b^{\prime}$ denote three wave-number intersection points for incident energy $E_K$ = 130 meV of electrons within range-II and $V_0$ = 50 meV or in situation-B of Fig. 1. (b) Polar plots of partial tunneling coefficients ${\mathcal{T}}_{s \rightarrow s^{\prime\prime}}$ as a function of incident angle $\theta_k$ for fixed $E_K$ = 130 meV and $V_0$ = 50 meV in N$_T$ = 2 different cases listed in Table\ \ref{tab-4}.}
\label{FIG5}
\end{figure}

For $V_0>0$ in Fig.\,\ref{FIG5}, we take a step-potential barrier height $V_0=50\,$meV and a kinetic energy $E_K=130\,$meV for spin-polarized incident electrons within the energy range II. From Fig.\,\ref{FIG5} and Table\ \ref{tab-4}, we realize that both inter-spin-channel ($c\to a'$ with spin procession) and intra-spin-channel ($c\to b'$ without spin procession) partial tunnelings of electrons appear in Cases-1 \& 2, respectively. However, no spin-procession process occurs for the reflection of electrons in these two cases. Furthermore, due to $k=k^\prime\geq q$ and $\theta_{\bf k}=\theta_{{\bf k}^\prime}\leq\theta^{\prime\prime}_{\bf q}$, a resonant partial intra-spin-channel tunneling in Case-2 suffers from an 
angle restriction beyond which tunneling coefficients ${\cal T}_{s\to s^{\prime\prime}}$ drops to zero. Meanwhile, for Case-1, on the other hand, we have $k=k^\prime\gg q$ and $\theta_{\bf k}=\theta_{{\bf k}^\prime}\ll\theta^{\prime\prime}_{\bf q}$, and then, only a relatively small and non-resonant partial inter-spin-channel tunneling 
with ${\cal T}_{s\to s^{\prime\prime}}<1$ is seen very close to $\theta_{\bf k}=0$, which is quite different from the observation of ${\cal T}_{s\to s^{\prime\prime}}\equiv 1$ for all $\theta_{\bf k}$ presented in Fig.\,\ref{FIG3} when $V_0=0$.
\medskip

\begin{table}[htbp]
\caption{\bf Configurations for $E_K=180\,$meV in range-III and $V_0=50\,$meV}
\begin{tabular}{ccccc}
\hline\hline
Case Number\ \ \ &  Incidence\ \ \ & Reflection\ \ \ & Transmission\ \ \ & Tunneling Status \\
\hline
$1$&  $d$&  $d$&  $c^{\prime}$&  $\checkmark$  \\
$2$&  $d$&  $e$&  $c^{\prime}$&  $\times$  \\
$3$&  $e$&  $d$&  $c^{\prime}$&  $\checkmark$  \\
$4$&  $e$&  $e$&  $c^{\prime}$&  $\checkmark$  \\
\hline\hline
\end{tabular}
\label{tab-5}
\end{table}
\medskip

\begin{figure} 
\centering
\includegraphics[width=0.55\textwidth]{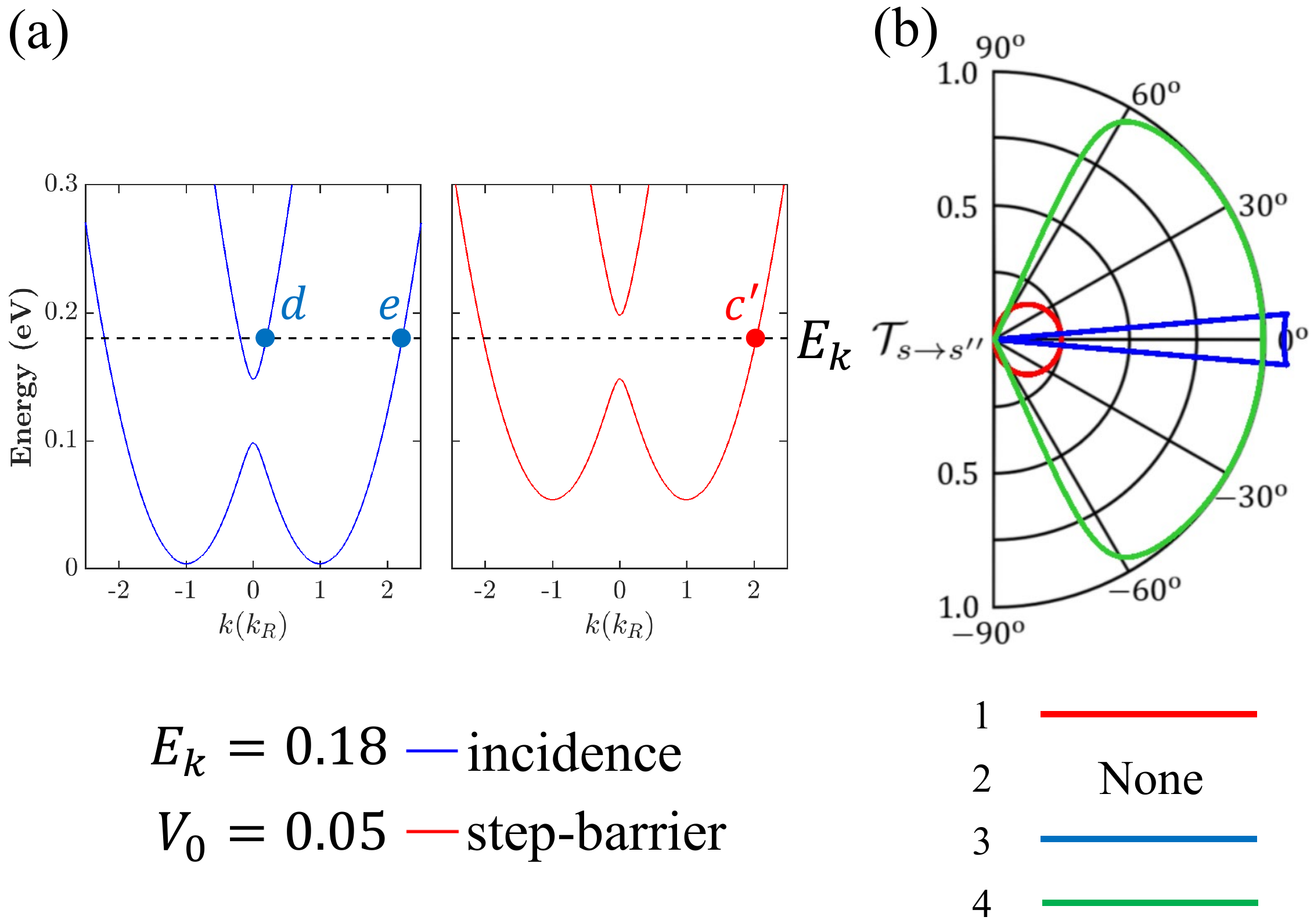}
\caption{(a) Calculated split Rashiba-Zeeman energy bands as a function of scaled electron wave number k/$k_R$, where the incident energy $E_K$ is indicated by a black-dashed line. Symbols $d$, $e$, $c^{\prime}$ denote three wave-number intersection points for incident energy $E_K$ = 180 meV of electrons within range-III and $V_0$ = 50 meV or in situation-B of Fig. 1. (b) Polar plots of partial tunneling coefficients ${\mathcal{T}}_{s \rightarrow s^{\prime\prime}}$ as a function of incident angle $\theta_k$ for fixed $E_K$ = 180 meV and $V_0$ = 50 meV in N$_T$ = 4 different cases listed in Table\ \ref{tab-5}.}
\label{FIG6}
\end{figure}

For the same barrier height $V_0=50\,$meV, numerical results are presented in Fig.\,\ref{FIG6} as we increase the incident energy $E_K=180\,$meV from range II to range III.
As found from Table\ \ref{tab-5}, for Case-1 we have $k=k^\prime\ll q$, and $\theta^{\prime\prime}_{\bf q}\ll\theta_{\bf k}=\theta_{{\bf k}'}$ as well due to $k_y=k_y'=q_y$. 
Then, we know from Eq.\,\eqref{boundary3} that $|t(k_x,k^\prime_x,q_x\,\vert s,s^\prime,s^{\prime\prime})|<1$ for partial inter-spin-channel tunneling of electrons. 
For Case-2, on the other hand, we get $k\ll q<k^\prime$ as well as $\theta_{\bf k}\gg\theta^{\prime\prime}_{\bf q}>\theta_{{\bf k}^\prime}$. In such a case, we find ${\cal T}_{s\to s^{\prime\prime}}/N_{\rm T}>1$ from Eq.\,\eqref{BigT}, and therefore, we expect that this partial inter-spin-channel tunneling of electrons should be excluded. Finally, for partial 
intra-spin-channel tunneling of electrons in Cases-3 \& 4, 
we have $k>q\gg k^\prime$ and $\theta_{\bf k}<\theta^{\prime\prime}_{\bf q}\ll\theta_{{\bf k}^\prime}$  
as well as $k=k^\prime>q$ and $\theta_{\bf k}=\theta_{{\bf k}^\prime}<\theta^{\prime\prime}_{\bf q}$. As a result, by using Eqs.\,\eqref{boundary3} and \eqref{BigT}, we reach the conclusions that  
$|t(k_x,k^\prime_x,q_x\,\vert s,s^\prime,s^{\prime\prime})|/\sqrt{N_{\rm T}}>1$ is disallowed while ${\cal T}_{s\to s^{\prime\prime}}\le 1$ is allowed for Cases-3 \& 4, separately, where both Case-3 and Case-4 are further limited by a critical angle $\theta_{\rm c}$ beyond which ${\cal T}_{s\to s^{\prime\prime}}$ reduces to zero.
\medskip

\begin{table}[htbp]
\caption{\bf Configurations for $E_K=180\,$meV in range-III and $V_0=100\,$meV}
\begin{tabular}{ccccc}
\hline\hline
Case Number\ \ \ &  Incidence\ \ \ & Reflection\ \ \ & Transmission\ \ \ & Tunneling Status \\
\hline
$1$&  $d$&  $d$&  $a^{\prime}$&  $\checkmark$  \\
$2$&  $d$&  $e$&  $a^{\prime}$&  $\checkmark$  \\
$3$&  $e$&  $d$&  $a^{\prime}$&  $\times$  \\
$4$&  $e$&  $e$&  $a^{\prime}$&  $\checkmark$  \\
$5$&  $d$&  $d$&  $b^{\prime}$&  $\checkmark$ \\
$6$&  $d$&  $e$&  $b^{\prime}$&  $\times$  \\
$7$&  $e$&  $d$&  $b^{\prime}$&  $\checkmark$  \\
$8$&  $e$&  $e$&  $b^{\prime}$&  $\checkmark$  \\
\hline\hline
\end{tabular}
\label{tab-6}
\end{table}
\medskip

\begin{figure} 
\centering
\includegraphics[width=0.65\textwidth]{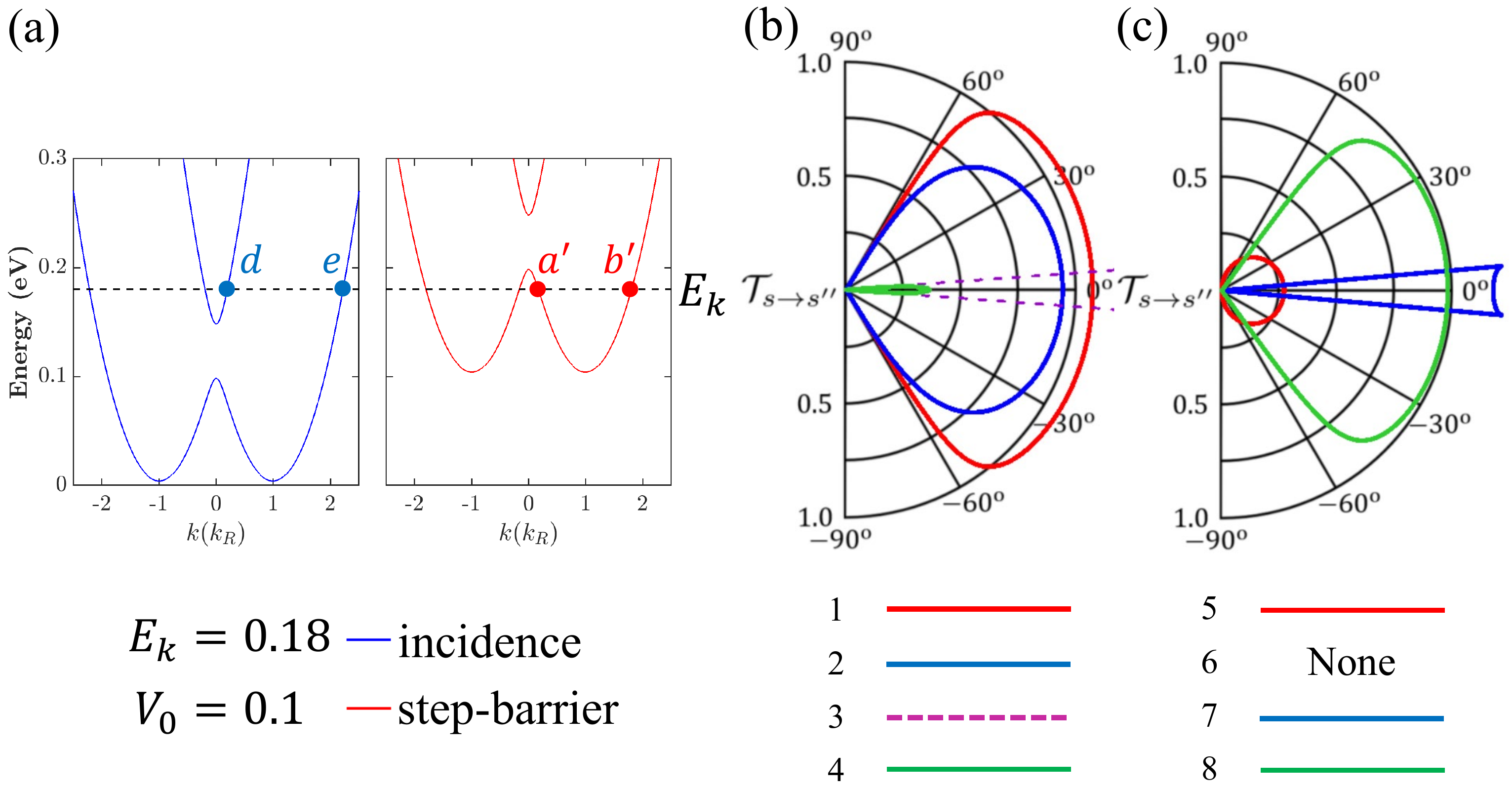}
\caption{(a) Calculated split Rashiba-Zeeman energy bands as a function of scaled electron wave number k/$k_R$, where the incident energy $E_K$ is indicated by a black-dashed line. Symbols $d$, $e$, $a^{\prime}$, $b^{\prime}$ denote four wave-number intersection points for incident energy $E_K$ = 180 meV and $V_0$ = 100 meV or in situation-D of Fig. 1. (b-c) Polar plots of partial tunneling coefficients ${\mathcal{T}}_{s \rightarrow s^{\prime\prime}}$ as a function of incident angle $\theta_k$ for fixed $E_K$ = 180 meV and $V_0$ = 100 meV in N$_T$ = 8 different cases listed in Table\ \ref{tab-6}.}
\label{FIG7}
\end{figure}

As the last situation considered in Sec.\,\ref{sec-4.2}, we would keep the incident kinetic energy $E_K=180\,$meV in range III while lift the barrier height $V_0$ from $50\,$meV to $100\,$meV, as seen in Fig.\,\ref{FIG7}. In this way, the kinetic energy of electrons on the barrier side moves down from range II in Fig.\,\ref{FIG6} into range I in Fig.\,\ref{FIG7}.
Compared with Fig.\,\ref{FIG6} and Table\ \ref{tab-5} with $N_{\rm T}=4$, we have found that the increase of $V_0$ in Fig.\,\ref{FIG7} has doubled the configuration space with $N_{\rm T}=8$ as listed in Table\ \ref{tab-6}. Here, we have 
$k\gg q\sim k^\prime$ or $\theta_{\bf k}\ll\theta^{\prime\prime}_{\bf q}\sim\theta_{{\bf k}^\prime}$ for Case-3, which leads to an excluded partial inter-spin-channel tunneling with $|t(k_x,k^\prime_x,q_x\,\vert s,s^\prime,s^{\prime\prime})|/\sqrt{N_{\rm T}}\gg 1$ 
according to Eq.\,\eqref{boundary3} 
and is further accompanied by a critical angle beyond which ${\cal T}_{s\to s^{\prime\prime}}$ drops to zero. Similarly, for Case-6, we find $k\ll q<k^\prime$ 
or $\theta_{\bf k}\gg\theta^{\prime\prime}_{\bf q}>\theta_{{\bf k}^\prime}$, and then we arrive at ${\cal T}_{s\to s^{\prime\prime}}/N_{\rm T}\gg 1$ by using Eq.\,\eqref{BigT} for an excluded 
partial inter-spin-channel tunneling.
On the other hand, we get $k=k^\prime>q$ or $\theta_{\bf k}=\theta_{{\bf k}^\prime}<\theta^{\prime\prime}_{\bf q}$ for Case-8. Therefore, from Eq.\,\eqref{BigT} we know the 
partial intra-spin-channel tunneling ${\cal T}_{s\to s^{\prime\prime}}\leq 1$ but it is still limited by a critical angle $\theta_{\rm c}$ for this case. A similar conclusion can be drawn for Case-1, where $k\sim q\sim k^\prime$ or 
$\theta_{\bf k}\sim\theta^{\prime\prime}_{\bf q}\sim\theta_{{\bf k}^\prime}$ which gives rise to the partial intra-spin-channel $|t(k_x,k^\prime_x,q_x\,\vert s,s^\prime,s^{\prime\prime})|>1$ based on Eq.\,\eqref{boundary3} but it is still allowed physically since $|t(k_x,k^\prime_x,q_x\,\vert s,s^\prime,s^{\prime\prime})|/\sqrt{N_{\rm T}}<1$. 
However, ${\cal T}_{s\to s^{\prime\prime}}$ in Case-1 reduces to zero once $\theta_{\bf k}$ becomes bigger than the critical angle $\theta_{\rm c}$ .
\medskip

For Cases-2 \& 5 in Fig.\,\ref{FIG7}, we acquire $k^\prime\gg q\sim k$ or $\theta_{{\bf k}^\prime}\ll\theta^{\prime\prime}_{\bf q}\sim\theta_{\bf k}$ 
and $k=k^\prime\ll q$ or $\theta_{\bf k}=\theta_{{\bf k}^\prime}\gg\theta^{\prime\prime}_{\bf q}$, respectively. As a result, we expect partial intra-spin-channel tunneling coefficient ${\cal T}_{s\to s^{\prime\prime}}<1$ from Eq.\,\eqref{BigT} 
for Case-3 while partial inter-spin-channel tunneling coefficient $|t(k_x,k^\prime_x,q_x\,\vert s,s^\prime,s^{\prime\prime})|\ll 1$ from Eq.\,\eqref{boundary3} for Case-5. For these two considered cases, 
angle restriction only applies to Case-2 but not to Case-5.
Finally, for Cases-4 \& 7, we have $k=k^\prime\gg q$ and $k^\prime\ll q<k$ separately, which corresponds to $\theta_{\bf k}=\theta_{{\bf k}^\prime}\ll\theta^{\prime\prime}_{\bf q}$ and 
$\theta_{\bf k}<\theta^{\prime\prime}_{\bf q}\ll\theta_{{\bf k}^\prime}$. Then, for the former we get partial inter-spin-channel tunneling coefficient ${\cal T}_{s\to s^{\prime\prime}}\ll 1$ 
from Eq.\eqref{BigT}, while we find partial intra-spin-channel tunneling coefficient ${\cal T}_{s\to s^{\prime\prime}}>1$ from the same Eq.\eqref{BigT} for the latter case,
where physical condition ${\cal T}_{s\to s^{\prime\prime}}/N_{\rm T}<1$ is satisfied. Moreover, an angle restriction $\theta_{\bf k}\leq\theta_{\rm c}$ has been further added to Case-7.
\medskip

Discussions on Figs.\,\ref{FIG5}-\ref{FIG7} are only limited to intra- and inter-spin-channel tunnelings of electrons. For cases of intra- and inter-spin-channel reflections of electrons, an analysis can be done in a similar way. 

\subsection{Tunneling Energy Spectra and Effect of Interface $\delta$-Function}
\label{sec-4.3}

\begin{figure} 
\centering
\includegraphics[width=0.55\textwidth]{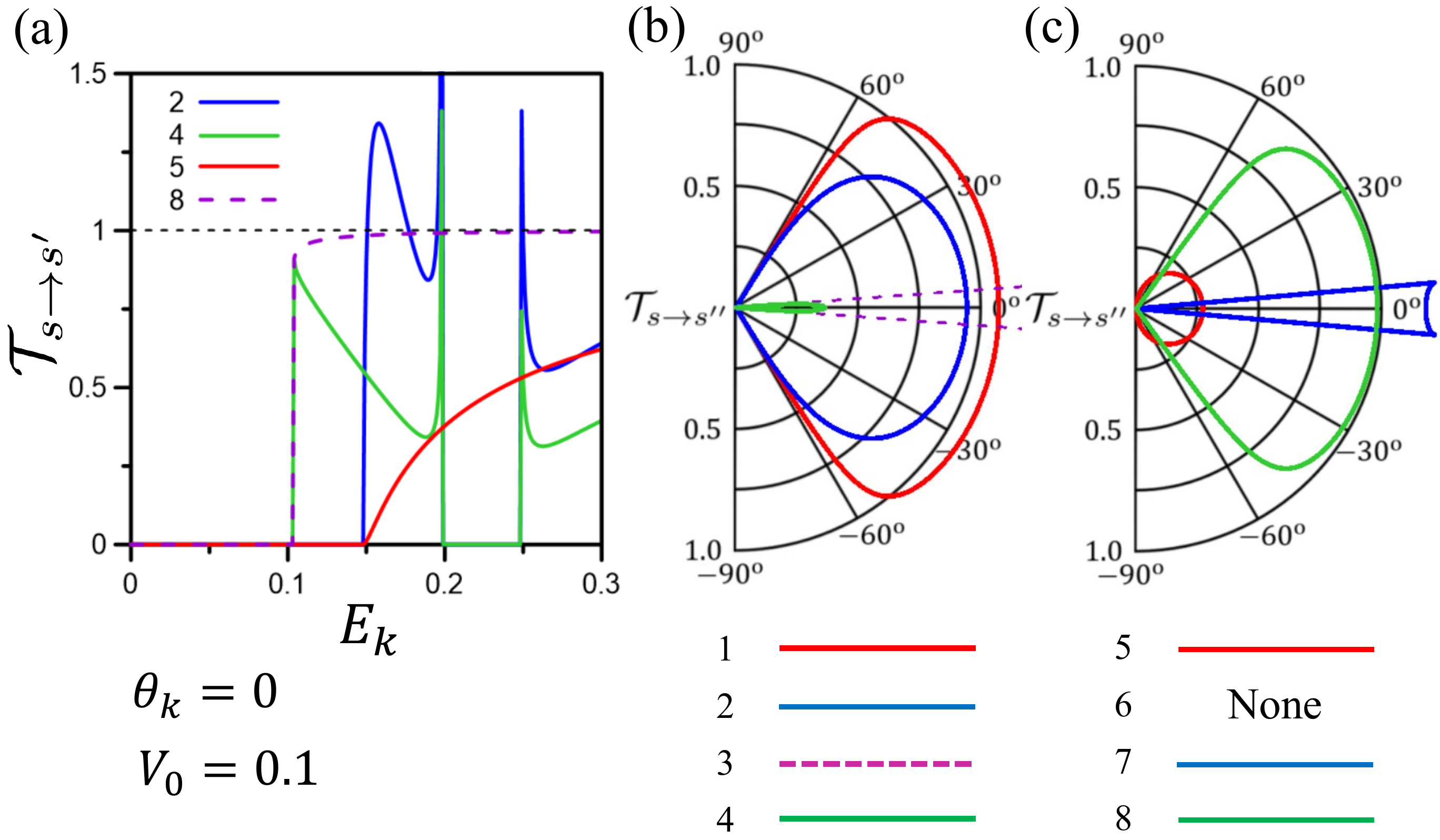}
\caption{(a) Calculated partial tunneling coefficients ${\mathcal{T}}_{s \rightarrow s^{\prime\prime}}$ as a function of the kinetic energy $E_K$ of incident electrons for $\theta_k$ = 0 and cases 3, 4, 5, 8 selected from Table VI. (b-c) Polar plots of partial tunneling coefficients ${\mathcal{T}}_{s \rightarrow s^{\prime\prime}}$, taken from Figs. 7(b) and 7(c) as a reference here, for fixed $E_K$ = 50 meV and $V_0$ = 100 meV in all cases listed in Table\ \ref{tab-6}.}
\label{FIG8}
\end{figure}

In Sec.\,\ref{sec-4.2}, we only deal with the incidence-angle dependence of partial intra- and inter-spin-channel tunneling coefficients ${\cal T}_{s\to s^{\prime\prime}}$. 
Here, we present in Fig.\,\ref{FIG8} the dependence of ${\cal T}_{s\to s^{\prime\prime}}$ on the kinetic energy $E_K$ of incident spin-polarized electrons at $\theta_{\bf k}=0$ and $V_0=50\,$meV. 
For specific, we select Cases-3,\,4,\,5,\,8 as listed in Table\ \ref{tab-6} for comparisons and discussions.
\medskip

For fixed $V_0=100\,$meV and $\theta_{\bf k}=0$ (or $k_y=0$), Cases-4 and 8 in Fig.\,\ref{FIG8} display a first threshold $E^{(1)}_{\rm T}=V_0=100\,$meV 
for tunneling coefficients ${\cal T}_{s\to s^{\prime\prime}}$, below which 
${\cal T}_{s\to s^{\prime\prime}}$ remains to be zero due to lack of available states for tunneling. Moreover, Cases-2 and 5 present a second threshold $E^{(2)}_{\rm T}=2\Delta_Z-E_{\rm min}$ for tunneling coefficients ${\cal T}_{s\to s^{\prime\prime}}$ since $2\Delta_Z-E_{\rm min}\sim 150\,\mbox{meV}>V_0$. As $E_K>E^{(1)}_{\rm T}$, ${\cal T}_{s\to s^{\prime\prime}}$ quickly approach unity for Case-8, 
while ${\cal T}_{s\to s^{\prime\prime}}$ for Case-4 further experiences a tunneling-forbidden region for $E_K$ within which ${\cal T}_{s\to s^{\prime\prime}}$ is suppressed to zero.
The presence of such a tunneling-forbidden region with respect to incident energy $E_K$ is related to the fact that $E_K$ enters into a Zeeman gap on the barrier side 
and then the tunneling-ending $a'$-state, illustrated in Fig.\,\ref{FIG7}, becomes inaccessible. However, such a tunneling-ending $a'$-state appears once again  as $E_K$ becomes above the 
Zeeman gap. As a result, the same tunneling-forbidden region is expected to show up for Case-2 based on the same reason. Finally, when $E_K$ goes beyond this tunneling-forbidden region, 
${\cal T}_{s\to s^{\prime\prime}}$ for all Cases-2,\,4,\,5,\,8 will gradually increase with $E_K$ until a unity value is reached.
\medskip

\begin{figure} 
\centering
\includegraphics[width=0.65\textwidth]{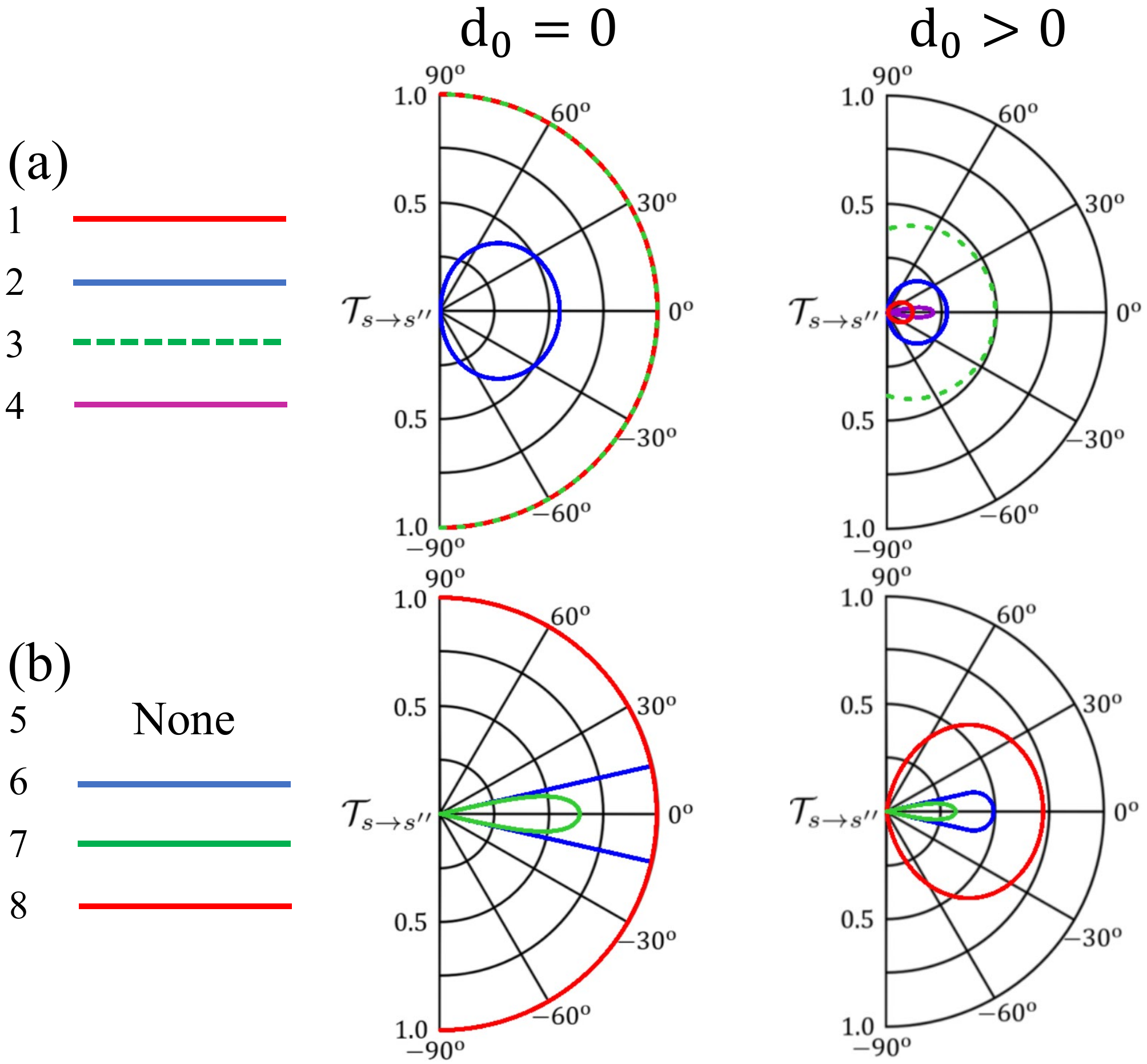}
\caption{Comparison of two calculated polar plots of partial tunneling coefficients ${\mathcal{T}}_{s \rightarrow s^{\prime\prime}}$ as seen in Figs. 2(b) and 2(c) with $d_0$ = 0 (Left) and with $d_0 >$ 0 (Right), respectively, for fixed $E_K$ = 50 meV and $V_0$ = 0, where all $N_T$ = 8 cases considered here are listed in Table\ \ref{tab-2}.}
\label{FIG9}
\end{figure}

Now, we turn to studying the effect of an interface $\delta$-function ($d_0>0$) on spin-procession processes in two different situations with either $V_0=0$ or $V_0>0$. 
From Eqs.\,\eqref{boundary5} and \eqref{boundary5p}, we know that one always obtain $|r|\neq 0$ and $|t|<1$ even in the case of $V_0=0$ if $d_0\neq 0$. 
By directly comparing two situations with ($d_0>0$) and without ($d_0=0$) an interface $\delta$-function in Fig.\,\ref{FIG9} and Fig.\,\ref{FIG2}, we find that previous Cases-4 and 5 
in Fig.\,\ref{FIG2} with no spin-procession processes change to opposite situations with weak but finite forward spin-procession processes between $b\to a'$ and $a\to b'$ states due to introduced discontinuity in their wave-function derivatives by a finite $d_0$ value, as can be verified from Eqs.\,\eqref{boundary4} and \eqref{boundary4p}. 
On the other hand, previous three Cases-1,\,3,\,8 without spin-procession process in Fig.\,\ref{FIG2} are greatly weakened in Fig.\,\ref{FIG9} for the first two but only slightly decreased for the last one. 
Meanwhile, the angle-restricted partial forward spin-procession process in Case-6 of Fig.\,\ref{FIG2} has been fully suppressed in Fig.\,\ref{FIG9}.
Finally, for the rest Cases-2 and 7 with partial spin-procession processes in Fig.\,\ref{FIG2}, the existence of an interface $\delta$-function in Fig.\,\ref{FIG9} 
seems to have no influence on situation in Case-2  but significantly broadens the critical angle $\theta_{\rm c}$ for Case-7.
\medskip

\begin{figure} 
\centering
\includegraphics[width=0.65\textwidth]{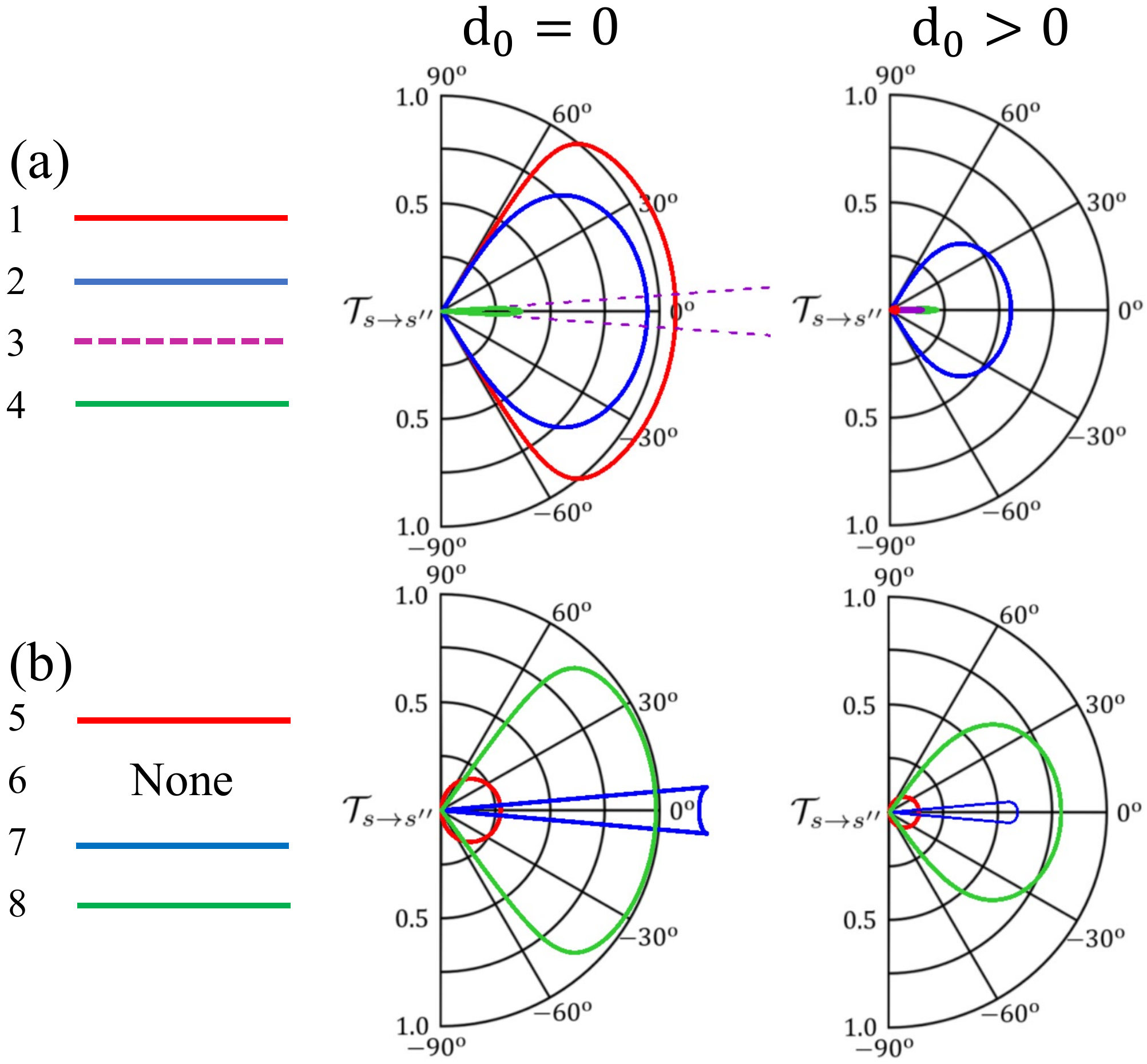}
\caption{Comparison of two calculated polar plots of partial tunneling coefficients ${\mathcal{T}}_{s \rightarrow s^{\prime\prime}}$ as seen in Figs. 7(b) and 7(c) with $d_0$ = 0 (Left) and with $d_0 >$ 0 (Right), respectively, for fixed $E_K$ = 180 meV and $V_0$ = 100 meV, where all $N_T$ = 8 cases considered here are listed in Table\ \ref{tab-6}.}
\label{FIG10}
\end{figure}

In order to reveal variation in intra- and inter-spin-channel tunnelings of electrons in Fig.\,\ref{FIG7} under an interface $\delta$-function, we compare calculated partial tunneling coefficients 
${\cal T}_{s\to s^{\prime\prime}}$ with $E_K=180\,$meV and $V_0=100\,$meV in Fig.\,\ref{FIG10} for all listed $N_{\rm T}=8$ cases in Table\ \ref{tab-6}. 
\medskip

Under an interface $\delta$-function, the previous excluded partial inter-spin-channel tunneling coefficients ${\cal T}_{s\to s^{\prime\prime}}/N_{\rm T}\gg 1$ in Case-6, as well as 
$|t(k_x,k^\prime_x,q_x\,\vert s,s^\prime,s^{\prime\prime})|/\sqrt{N_{\rm T}}\gg 1$ in Case-3, of Fig.\,\ref{FIG7} switch to two weakened allowed ones in Fig.\,\ref{FIG10} with and without an angle restriction, 
respectively. On the contrary, an excluded partial inter-spin-channel tunneling coefficient ${\cal T}_{s\to s^{\prime\prime}}/N_{\rm T}\gg 1$ in Case-5 of Fig.\,\ref{FIG10} corresponds to an allowed one in Fig.\,\ref{FIG7} 
with no angle restriction. 
Additionally, previous full angle-restricted partial intra-spin-channel tunneling coefficients ${\cal T}_{s\to s^{\prime\prime}}$ for Cases-8 \& 1 of Fig.\,\ref{FIG7} decrease greatly in Fig.\,\ref{FIG10} 
free with an angle restriction. Furthermore, under the same interface $\delta$-function, 
the angle-restricted partial intra-spin-channel tunneling coefficient ${\cal T}_{s\to s^{\prime\prime}}$ for Case-2 of Fig.\,\ref{FIG7} changes into a quite extraordinary one 
acquiring ${\cal T}_{s\to s^{\prime\prime}}\neq 0$ in Fig.\,\ref{FIG10} even at $\theta_{\bf k}=\pm\pi/2$.
Finally, for Cases-4 \& 7, the sharply angle-restricted partial inter-spin-channel tunneling coefficient ${\cal T}_{s\to s^{\prime\prime}}$ for the former in Fig.\,\ref{FIG7} remains largely unchanged in Fig.\,\ref{FIG10},
however the angle-restricted intra-spin-channel tunneling coefficient ${\cal T}_{s\to s^{\prime\prime}}$ for the latter in Fig.\,\ref{FIG7} is reduced significantly in Fig.\,\ref{FIG10}.

\subsection{Quantum Interference of Correlated-Pair of Spin States of Electrons}
\label{sec-4.4}

In Sec.\ \ref{sec-4.4}, we would utilize the calculated transmission amplitude $t_{1,2}(k_x,k^\prime_x,q_x\,\vert s,s^\prime,s_{1,2}^{\prime\prime};d_0)$ and reflection amplitude $r_{1,2}(k_x,k^\prime_x,q_x\,\vert s,s_{1,2}^\prime,s^{\prime\prime};d_0)$ from Eqs.\,\eqref{boundary5} and \eqref{boundary5p}, and then apply them to Eqs.\,\eqref{new-11} and \eqref{new-12} for calculating the total tunneling probability density $|\Psi^{({\rm t})}_{\rm tot}(\mbox{\boldmath$r$})|^2$ 
and the total reflection probability density $|\Psi^{({\rm r})}_{\rm tot}(\mbox{\boldmath$r$})|^2$. In this way, we are able to quantify the quantum-interference effects between pair of correlated spin states of electrons on either tunneling or reflection side of an interface in the system, respectively.
\medskip

\begin{table}[htbp]
\caption{\bf Configuration for $E_K=180\,$meV in range-III and $V_0=100\,$meV}
\begin{tabular}{cccc}
\hline\hline
Incidence\ \ \ & Transmission\ \ \ & Reflection-1\ \ \ & Reflection-2 \\
\hline
$d$&  $a^{\prime}$&  $d$&  $e$  \\
\hline\hline
\end{tabular}
\label{tab-7}
\end{table}
\medskip

\begin{figure} 
\centering
\includegraphics[width=0.55\textwidth]{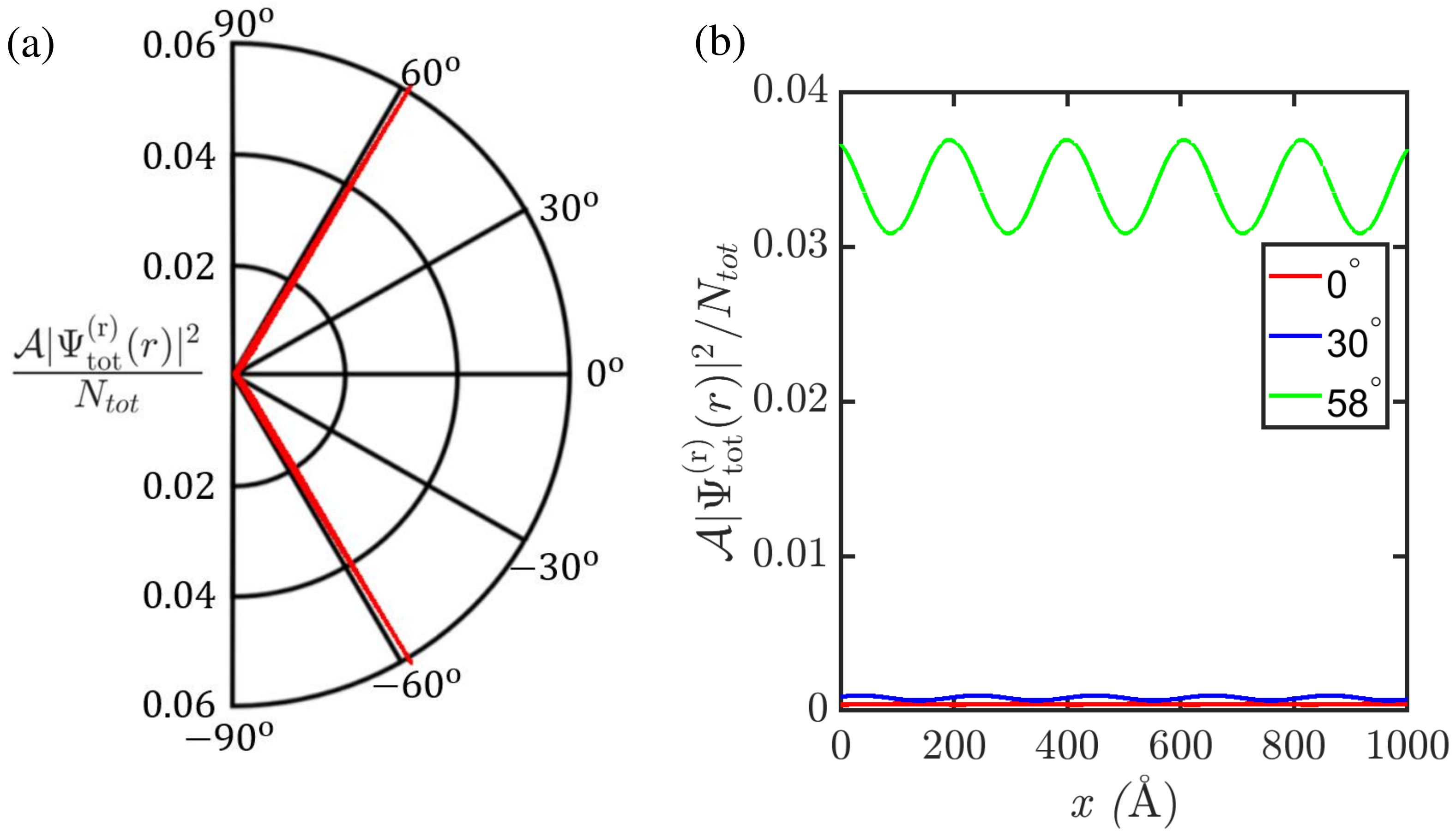}
\caption{Plots for total reflection-wavefunction density ${\cal A}|\Psi^{({\rm r})}_{\rm tot}(\mbox{\boldmath$r$})|^2/N_T$ in Eq.\,\eqref{new-12} at the interface $x=0$ as a function of incident angle $\theta_{\bf k}$ in $(a)$ as well as a function of position $x$ on the reflection side in $(b)$ for three different incident angles $\theta_{\bf k}=0^{\rm o},\,30^{\rm o},\,58^{\rm o}$. Here, we set $E_K$ = 180 meV, $V_0$ = 100 meV and $N_T$ = 8 for the case listed in Table\ \ref{tab-7}.}	
\label{FIG11}
\end{figure}

For this purpose, let us first study the case listed in Table\ \ref{tab-7}, corresponding to cases 1 \& 2 in Table\ \ref{tab-6}, for a pair of correlated spin states ($d$ \& $e$) of electrons on the reflection side under a single incident spin state $d$ of electron, i.e., $d\to d$ (without a spin flip) and $d\to e$ (with a spin flip) in the absence of an interface $\delta$-function or $d_0=0$.
\medskip

\begin{table}[htbp]
\caption{\bf Configuration for $E_K=180\,$meV in range-III and $V_0=100\,$meV}
\begin{tabular}{cccc}
\hline\hline
Incidence\ \ \ & Reflection\ \ \ & Transmission-1\ \ \ & Transmission-2 \\
\hline
$d$&  $d$&  $a^{\prime}$&  $b^{\prime}$  \\
\hline\hline
\end{tabular}
\label{tab-8}
\end{table}
\medskip

From results in Fig.\,\ref{FIG11}, we know that whenever the incident angle $\theta_{\bf k}$ becomes far away from its angular-distribution boundaries at $\pm 60^{\rm o}$, 
the superposition of a pair of correlated electron spin states $d$ \& $e$ on the reflection side presents no sign of quantum interference 
and also becomes very small due to their similar amplitudes and nearly out-of-phase cancellation between them, as can be easily verified from the right panel of Fig.\,\ref{FIG11}. 
However, as $\theta_{\bf k}=58^{\rm o}$, the quantum-interference effect is greatly enhanced and gives rise to a periodically oscillating function of the position $x$ on the reflection side away from the interface at $x=0$ 
with a spatial period around $200\,$\AA. Physically, such a quantum interference results from the unique band structure associated with Rashba spin-orbital coupling, and meanwhile provides us with a tool for 
controlling reflection of an incident electron with a fixed kinetic energy $E_K$ and a step-barrier height $V_0$ with or without a spin flip. 
\medskip

Next, we consider the case in Table\ \ref{tab-8}, which is associated with cases 1 \& 5 in Table\ \ref{tab-6}, for another pair of correlated spin states ($a'$ \& $b'$) of electrons on the tunneling side 
under a single incident spin state $d$, i.e., $d\to a^\prime$ (without a spin flip) and $d\to b^\prime$ (with a spin flip) under $d_0=0$. 
\medskip

\begin{figure} 
\centering
\includegraphics[width=0.55\textwidth]{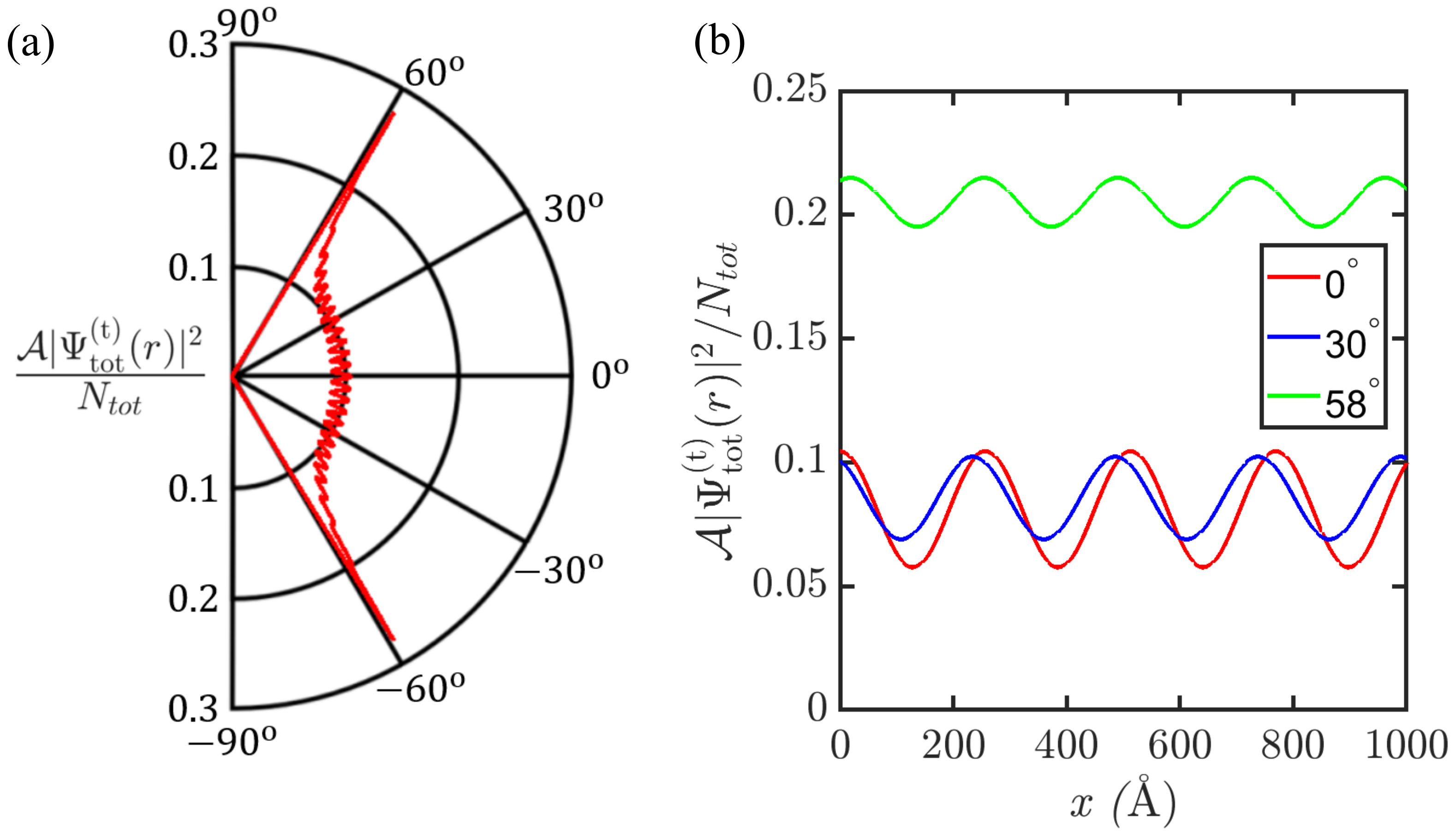}
\caption{Plots for total tunneling-wavefunction density ${\cal A}|\Psi^{({\rm t})}_{\rm tot}(\mbox{\boldmath$r$})|^2/N_T$ in Eq.\,\eqref{new-11} at the interface $x=0$ as a function of incident angle $\theta_{\bf k}$ in $(a)$ as well as a function of position $x$ on the tunneling side in $(b)$ for three different incident angles $\theta_{\bf k}=0^{\rm o},\,30^{\rm o},\,58^{\rm o}$. Here, we set $E_K$ = 180 meV, $V_0$ = 100 meV and $N_T$ = 8 for the case listed in Table\ \ref{tab-8}.}	
\label{FIG12}
\end{figure}

From results presented in Fig.\,\ref{FIG12}, we observe that the superposition of a pair of correlated electron spin states $a'$ \& $b'$ on the tunneling side leads to a strong quantum interference between them 
and, in contrast to those in Fig.\,\ref{FIG11}, no cancellation between them are found, as seen clearly from the right panel of Fig.\,\ref{FIG12}. 
Specifically, for $\theta_{\bf k}=0^{\rm o}$ and $30^{\rm o}$, the resulting interference patterns are strong, similar to each other but with a relative phase sift between them, and 
acquire an enlarged oscillation period around $300\,$\AA. 
For $\theta_{\bf k}=58^{\rm o}$, on the other hand, the spatially-oscillating feature is found slightly weakened as a function of the position $x$ on the tunneling side away from the interface at $x=0$ 
although the total amplitude is increased due to nearly in-phase superposition of two correlated electron spin states $a'$ \& $b'$. 
Furthermore, we also find from the left panel of Fig.\,\ref{FIG12} that a quantum-interference pattern within a limited angle region $|\theta_{\bf k}|<60^{\rm o}$ 
occurs even at the interface $x=0$ due to different diffraction angles $\theta^{\prime\prime}_{{\bf q}_{1,2}}$ 
for electron tunneling as a function of incident angle $\theta_{\bf k}$, instead of position $x$ dependence 
in the right panel of Fig.\,\ref{FIG12}. 

\section{Summary}
\label{sec5}

In conclusion, we have established a theory for calculating the transmission $\mbox{\boldmath$J$}^{({\rm t})}(s'',\mbox{\boldmath$q$};\,|t|^2)$ 
and reflection $\mbox{\boldmath$J$}^{({\rm r})}(s',\mbox{\boldmath$k$}';\,|r|^2)$ probability currents 
of a charged particle across a steepness-enhanced potential step $V_B(x)=V_0\Theta(x)+d_0\delta(x)$ within a quantum well in the presence of a Rashba spin-orbit interaction. 
By using calculated energy eigenstate $E_s(k)$ for Rashba-Zeeman coupled two-dimensional conduction electrons in the presence of a spin-split gap $\Delta_Z$, 
both reflection and transmitted probability currents are explicitly computed  
for electrons under the potential step $V_B(x)$ within a quantum well. Moreover, by matching the spin-dependent boundary conditions at the interface $x=0$ for a barrier step, both reflection $r(k_x,k'_x,q_x\,\vert s,s',s'';d_0)$ 
and transmission $t(k_x,k'_x,q_x\,\vert s,s',s'';d_0)$ coefficients are determined 
as functions of incident kinetic energy $E_K$, selected incident-electron spin stets $s=\pm 1$ and angle of incidence $\theta_{\bf k}$ for different step-barrier heights $V_0$ and steepness $d_0$, 
as well as of various spin directions $s',\,s''=\pm 1$ for reflection and tunneling of electrons, Zeeman gaps $\Delta_Z$, and Rashba parameter values $\alpha_R$.
\medskip

In particular, our model system contains multiple spin channels for both reflection and tunneling of an incident electron. For this situation, we have introduced average transmission 
${\cal T}_{\rm av}(\mbox{\boldmath$k$},\mbox{\boldmath$k$}',\mbox{\boldmath$q$})$ and reflection ${\cal R}_{\rm av}(\mbox{\boldmath$k$},\mbox{\boldmath$k$}',\mbox{\boldmath$q$})$ coefficients by means of 
the total number $N_T$ of inequivalent cases for reflection or tunneling so that ${\cal T}_{\rm av}(\mbox{\boldmath$k$},\mbox{\boldmath$k$}',\mbox{\boldmath$q$})+{\cal R}_{\rm av}(\mbox{\boldmath$k$},\mbox{\boldmath$k$}',\mbox{\boldmath$q$})\equiv 1$ 
can be maintained.  
By varying $E_K$ and $s=\pm 1$ for incident electrons or $V_0$ and $d_0$, different features in tunneling and reflection of electrons have been demonstrated 
with the inclusion of inter-channel electron tunneling and reflection processes. Importantly, these spin-dependent unique features are further accompanied by spin-state quantum interference for 
either reflected or transmitted two spin-correlated electrons with the same $E_K$ but unequal reflection $\theta_{{\bf k}'_{1,2}}$ or diffraction $\theta^{\prime\prime}_{{\bf q}_{1,2}}$ angles. 
The distinctive property predicted in this paper is expected to acquire a lot of applications in both non-magnetic spintronics and quantum-computation devices.

\clearpage
\section*{Acknowledgments}

P.-H. Shih would like to thank the Ministry of Science and Technology of Taiwan for the support through Grant No. MOST 110-2636-M-006-002. 
G.G. would like to acknowledge the support from the Air Force Research Laboratory (AFRL) through Grant No. FA9453-21-1-0046.
D.H. would like to thank the Air Force Office of Scientific Research (AFOSR) for support.

\end{document}